\documentclass{aa} 
\usepackage{graphicx}
\usepackage{txfonts}
\usepackage{lipsum}
\usepackage{subcaption}         
\usepackage{lscape}             
\usepackage{placeins}           

\usepackage{xcolor}
\usepackage[dvipsnames]{xcolor}   

\begin{document}

   \title{Searching for core-collapse supernovae in binaries with ZTF}


%

   \author{Sylvia J. Zhu\inst{1}\fnmsep\thanks{Corresponding author: sylvia.zhu@desy.de}
        \and Jakob Nordin\inst{2}
        \and Emma de O\~{n}a-Wilhelmi\inst{1}
        \and Andrea Ercolino\inst{3}
        }

   \institute{Deutsches Elektronen-Synchrotron (DESY), Platanenallee 6, D-15738 Zeuthen, Germany
   \and Institut f\"{u}r Physik, Humboldt-Universit\"{a}t zu Berlin, D-12489 Berlin, Germany
   \and Argelander Institut f\"{u}r Astronomie, Auf dem H\"{u}gel 71, DE-53121 Bonn, Germany}

   \date{Received August 3, 2026}

 
  \abstract
   {
   Stripped envelope supernovae occur in massive stars that have lost their outer layers
   before exploding, possibly from binary interactions.
   Recently, SN~2022jli was
   detected with significant periodic oscillations in its lightcurve, likely
   due to accretion from a binary companion onto the newborn compact object.}
   {While evidence for periodic oscillations has been found in a few bright
   and noteworthy supernovae, the population properties are best understood with
   systematic searches of large data sets. We present prospects for running
   a search on data from ZTF and similar surveys that is flexible and
   computationally inexpensive, with the goal of setting up a search in realtime
   on data from ZTF, LSST, and similar surveys.}
   {We simulate core collapse supernova lightcurves with additional features
   from the binary interactions. We use spline fits to remove the underlying
   supernova behavior and isolate the periodicities, then apply a Lomb-Scargle
   periodogram to recover the simulated period. By varying the conditions,
   we can make recommendations for different kinds of data sets. Finally, we
   run the search on two sets of archival ZTF supernovae to illustrate the concept.}
    {
    For a wide range 
    of binary interaction parameters, we find that a long baseline (200 days) of
    observations with a three-day cadence effectively retrieve
    $\approx$\,50\% of the binary
    oscillations.
    However, this rate plummets with more typical 
    observation durations; 
    only a few percent are recoverable with a 30-day observation,
    and 10\% with a 75-day observation. 
    A substantial fraction of supernovae could therefore have 
    binary companion interactions that have simply been undetected. 
    Out of 212 moderately bright and well observed ZTF supernovae, we find suggestions
    of periodic oscillations in two candidates.
    }
   {}


   \maketitle

\nolinenumbers

\section{Introduction}

Stars with masses $\gtrsim\!10\,\mathrm{M}_\odot$ end their lives as 
core collapse (CC) supernovae (SNe). Based on their spectra,
they are categorized as a subtype of
either type I (no hydrogen lines) or type II (hydrogen lines) SNe. 
The CC SNe classified as type I
are also referred to as stripped envelope SNe, along with the 
type IIb (weak hydrogen lines), as the lack of
hydrogen and sometimes helium lines indicates that the outer layer has
been depleted before explosion \citep{Filippenko_1997}.
The progenitors of stripped envelope SNe
could be single massive stars with strong stellar winds,
or massive stars in binaries that have lost their envelopes due to 
binary interactions \citep{Conti_1975, Conti_1996, Podsiadlowski+_1992} \citep[see also][for a recent review]{Jerkstrand+_2026}.
In fact, the majority of stars 
with masses above a few M$_\odot$ are known to exist in multiple star 
systems (e.g., Fig.~1 of \citet{Offner+_2023}), and simulations
indicate that over half of stripped envelope SNe come from 
massive stars in binary systems \citep{Kochanek_2009, Zapartas+_2017, Zapartas+_2026}.
Late-time images of CC SNe have uncovered five binary companions
to date
\citep[and references therein]{Zapartas+_2026}.

Some of the SN themselves should also show evidence of binary interactions.
This was seen in SN~2022jli,
a type Ic at a distance of 23~Mpc \citep{SN2022jli_Moore+,
SN2022jli_Chen+}: After the initial decline, the SN then
rebrightened approximately 50~days after detection
and reached a secondary peak as bright as the initial brightness
at detection. It also showed strong periodic oscillations
with a period of 12~days during
this rebrightening episode, which lasted for around 200~days. 
In the binary scenario, the
SN shock disturbs the companion's outer layers and causes
them to inflate. The newborn compact object then accretes this
inflated material, resulting in both the rebrightening
and periodic oscillations 
\citep{Hirai+_2025, Lu+_arXiv_2025, King+Lasota_2024}.
In addition, shifts in the
velocities of the H$\alpha$ \citep{SN2022jli_Chen+} and Pa$\beta$
\citep{Cartier+_2026} lines were discovered that were
consistent with the orbital modulation deduced from the 
periodic oscillations. This all paints a picture of accretion
from a binary companion as opposed to interactions between the
supernova shock and overdensities in the circumstellar environment, 
an alternative explanation for short-term
variability in SN lightcurves 
\citep[e.g.,][]{Nagao+_2026}. The nature of the newborn
compact object is unknown, although the supernova brightness
and presumed strength of the natal kick suggest a neutron star
is more likely than a black hole. 
In the case of a neutron star, a magnetar
engine would provide enough energy to explain the rebrightening
but not the lightcurve oscillations or the hydrogen line velocity shifts 
\citep{Orellana+_2025, Cartier+_2026}. Additionally, evidence for
a gamma-ray excess at 1~GeV starting half a year after the initial
SN was reported by \citet{SN2022jli_Chen+} and \citet{SN2022jli_Zhang+},
wit the gamma-ray arrival times being potentially consistent with the
12.4-day periodicity.
Any gamma-ray excess could be due to either an accretion-powered
relativistic jet or a magnetar central engine \citep{Hirai+_2025},
and the six-month delay caused by the region's initially high
opacity to gamma rays.

A few other candidates for periodic lightcurve
oscillations have been reported. Most prominently, a 32-day period
was found in SN~2022esa \citep{SN2022esa}, a type Ic-CSM at a distance
of around 100~Mpc. For this SN, the oscillations began 
immediately after the start of the main SN peak, and there was no
rebrightening episode. The authors interpret the 
data as being more consistent with 
circumstellar material interaction, although they do not rule out a binary companion
origin. An 8.4-day periodicity
was also reported for SN~2015ap \citep{SN2015ap_Ragosta+}.
The observed lightcurve
modulations in SN~2022jli, SN2022esa, and SN2015ap may arise from
the interaction between a newly born compact object and a 
companion star \citep{Hirai+_2025, Ercolino+2026_arxiv} while
in other transients, especially superluminous supernovae, these
undulations might instead be related to 
precession effects from a magnetar central engine
\citep{Mashhoon+_1984, Hosseinzadeh+_2022, Farah+_2026}.

Massive stars in binaries tend to explode at particularly 
small orbital separations from their companions \citep{Sana+2012}.
If these binaries remain bound after the SN explosion, they 
could evolve into a class of accreting binary 
\citep{King_1988, Shao+Li_2015}. If the companion is also sufficiently
massive to undergo a supernova explosion, the 
compact objects could eventually
merge to produce the gravitational-wave signals that have been
detected by LIGO, Virgo, and KAGRA \citep{GWTC5}. 
Observations of more SN~2022jli-like systems would 
shed light on the evolution
of massive stars in binaries. However, the few such SNe found so far
have been particularly bright and/or well observed, and so are not
necessarily representative of the average SN. 

To better understand the population of such systems, it is important to search for
evidence of them systematically. We present prospects
for systematically searching for periodicities in CC SNe lightcurves,
using the Zwicky Transient Facility \citep[ZTF,][]{ztf} 
as our reference observatory. 
In our simulations, we focus on simulating lightcurves that would 
cover a large parameter space of observational properties.
We explore how to set up a broad search to identify
potentially interesting candidates, which could then be followed up
with further observations and/or examined with a more tailored approach.
Our findings can be used to inform search setups under more general 
conditions.

We describe our choices for the simulated SN lightcurves and binary interaction
parameters in Sec.~\ref{sec:simulations} and the search procedure development,
such as the choice of spline fit and periodogram grid, in Sec.~\ref{sec:development}.
We then discuss some results when applying various realistic observing
conditions in Sec.~\ref{sec:results}. In Sec.~\ref{sec:realdata} we run
our search on two sets of ZTF SNe as well as a few known SNe with
periodic oscillations. Finally, we discuss our findings and prospects for
future searches in Sec.~\ref{sec:conclusions}.

\section{Simulations}
\label{sec:simulations}

For the supernova lightcurves,
we used the \texttt{realize\_lcs()} function in \texttt{sncosmo} \citep{sncosmo}. We simulated 10 000 SN lightcurves with 
additional binary-induced features
to characterize the potential search
sensitivity to a wide range of binary parameters. After simulating
the lightcurves, we applied a spline fit to subtract out the
underlying supernova behavior and a Lomb-Scargle periodogram on the
residuals to isolate periodic signals, which will be discussed
fully in Sec.~\ref{sec:development}.

We used SN 2022jli
as a guideline but did not adhere too tightly to its properties to 
avoid biasing our simulations with a single example.
For reference, SN 2022jli had a peak $g$-band magnitude of around 15, 
equivalent to a scaled flux of $10^4$ assuming a zeropoint magnitude 
$\mathrm{ZP} = 25$ (Sec.~\ref{sec:basics}).
The oscillations had a period of 12.4~days and
an amplitude of around 10\% in flux. The SN 
rebrightened 20~days after detection.
The oscillations were significantly detected only after the 
rebrightening peak \citep{SN2022jli_Moore+, SN2022jli_Chen+}.

We performed the simulations and searches in the ZTF $g$ and $r$ bands separately,
as the relative cadence of observations in the
two bands varies, and combining the results can be nontrivial if
the color is changing. In the following sections, we focus
on the $g$-band results and report $r$-band results when they
differ.

\subsection{Supernovae}

We generated the base SNe with a wide variety of SN templates.
While the generalized version of the Lomb-Scargle
periodogram can account for a non-zero mean, it still assumes
a constant offset and cannot account for more complex behavior. 
The range of parameters is listed in Table~\ref{tab:params}.

\begin{table}[ht!]
\begin{center}
\begin{tabular}{|c|c|}
     \hline
     parameter & range \\
     \hline
     SN $z$ & [0.001, 0.1] \\
     SN scaling & [5$\times10^{-15}, 5\times10^{-14}$] \\
     \hline
     $T_\mathrm{osc}$ (d) & [1, 100] \\
     $A_\mathrm{osc}$ (ZP = 25) & [20, 200] \\
     $\Delta_\mathrm{osc}$ (d) & [0, 100] \\
     \hline
     $F_\mathrm{rebr}^\mathrm{max}$ (ZP = 25) & [0, 1000] \\
     $\tau_r$ (d) & [5, 20] \\
     $\tau_d$ (d) & [20, 50] \\
     $n$ & [1, 4] \\
     $\Delta_\mathrm{rebr}$ (d) & [10, 30] \\
     \hline
\end{tabular}
\caption{Parameter ranges as inputs to the lightcurve simulations. The subscript ``osc''
refers to the periodic oscillations and ``rebr'' to the
rebrightening features. The SN $z$ is used for redshifting
the observed wavelengths and time dilating the lightcurves,
while the dimensionless scaling parameter controls the brightness.
The oscillations are defined by their period $T_\mathrm{osc}$, 
amplitude $A_\mathrm{osc}$, and delay from the SN $t_0$ $\Delta_\mathrm{osc}$.
The rebrightening parameters are defined in Eq.~\ref{eq:rebrightening}.
All parameters are drawn uniformly from their respective ranges except for the
SN scaling parameter, which is drawn log-uniformly.
The parameters $A_\mathrm{osc}$ and 
$F_\mathrm{rebr}^\mathrm{max}$ are values of scaled flux, 
corresponding to $\mathrm{ZP}=25$ (Sec.~\ref{sec:basics}).}
\label{tab:params}
\end{center}
\end{table}

To simulate the base SNe, we used \texttt{sncosmo v2.12.1} \citep{sncosmo}
and uniformly sampled from the 115 built-in templates associated with 
CC SNe (after excluding a few that
returned empty lightcurves in the ZTF $g$ and $r$ bands).
While it is likely that massive stars in close binaries 
would preclude
hydrogen-rich SNe, we included all CC SNe templates to test our method's
robustness to a wider range of SN behavior. In fact, we found less
than a few percent improvement in sensitivity when only using templates for
stripped envelope supernovae (Sec.~\ref{app:SE}), suggesting that our
method is indeed able to well account for variations in supernova lightcurves.
For each SN, we chose the
scaling parameter such that the lightcurve maxima would range from
20 to 14 mag, with a peak in the distribution between 18 and 17 mag.
We drew the scaling parameter from a log uniform 
distribution for simplicity.
Note that $z$ 
redshifts the energies and dilates the timescales,
but does not affect the brightness scaling via the
distance,
so t maximum $z$ does not represent a detection horizon.

Using the realize\_lcs() function of \texttt{sncosmo}, we simulated
lightcurves in the ZTF $g$- and $r$-bands with values of skynoise
$\sigma_\mathrm{sky}$ drawn from fits to ZTF data
(Sec.~\ref{sec:skynoise}). 
Since real observations tend to deviate
from a strictly regular cadence,
we added an offset in time drawn from a normal distribution
with a standard deviation of 2~h.
We simulated the lightcurves out to 200 days
after $t_0$, where $t_0$ is 
usually the maximum of the SN peak.

\subsection{Binary interaction properties}

We selected the period $T_\mathrm{osc}$
to have a minimum of 1~d and a maximum of 100~d. These limits
are longer than the periods of known X-ray binaries, which range
from sub-hours to tens of days, and which are
the later evolutionary stages of our systems. Indeed, 
\citet{Ercolino+2026_arxiv} found post-SN binary 
orbital periods of
$\mathcal{O}$(10~d) using a population synthesis framework, although  
periods outside of this range are also possible depending on the
population assumptions (Fig.~7 in the referenced work).

The oscillations detected in SN~2022jli were skewed rather than
symmetric \citep{SN2022jli_Moore+, SN2022jli_Chen+},
likely due to the orbital eccentricity 
caused by the compact object's natal kick \citep{Hirai+_2025}. 
This could be common for these systems,
as eccentric orbits allow for closer periastrons and therefore a larger amount of accretable matter
\citep{Ercolino+2026_arxiv}. On the other hand, the oscillations in 
SN~2022esa were
symmetric \citep{SN2022esa}; if these are 
also due to binary accretion, then skewness is not necessarily
universal. For simplicity, 
we mainly simulate the accretion-induced oscillations 
as a pure sinusoid. We also explored
the effect of using skewed sawtooth oscillations, and found that the search sensitivity only decreases
by a few percent (Sec.~\ref{app:sawtooth}).

We set the amplitude of the oscillations $A_\mathrm{osc}$ 
to range from below the level of noise (scaled flux of 20;
see Eq.~\ref{eq:skynoise}) to
bright enough to be visible by eye (scaled flux of 200, 
equivalent to $\Delta m = 0.05$~mag at 16 mag or 
$\Delta m = 0.3$~mag at 18 mag). This maximum
was chosen as the point where the performance
began to plateau, so that including brighter 
oscillations would improve the signal recovery performance but not reveal
new information. We assume that $A_\mathrm{osc}$ does 
not change over the observation duration.
To zeroth order, we would expect the accretable
material to deplete over time, so our simulated signals
are optimistic in this respect. We also assume the same
$A_\mathrm{osc}$ in the $g$ and $r$ bands. Note that
\citet{Cartier+_2026} found a larger oscillation amplitude
for SN~2022jli in the bluer bands; however, since we
are considering the two sets of simulations separately,
this assumption should not have an effect on our results.
We set the oscillations to begin a random number of days
$\Delta_\mathrm{osc}$ after $t_0$,
as the oscillations in SN~2022jli and SN~2022esa had different
delay times.

We simulated rebrightening
features using a function that describes a fast rise and
exponential decay:
\begin{align}\label{eq:rebrightening}
    F_\mathrm{rebr}(t) = A_\mathrm{rebr}\left(1-e^{-t/\tau_r}\right)^n e^{-t/\tau_d}
\end{align}
where $t$ is the number of days since $t_0$, $A_\mathrm{rebr}$ is
the amplitude in scaled flux, $\tau_r$ and $\tau_d$ are
the rise and decay timescales in days, respectively,
and $n$ moderates the sharpness of the peak with larger
values of $n$ producing broader peaks.
In practice, 
we set the maximum
amplitude of the rebrightening feature $F^\mathrm{max}_\mathrm{rebr}$
and then calculated $A_\mathrm{rebr}$ from Eq.~\ref{eq:rebrightening}.
We set the rebrightening to begin a number of days
$\Delta_\mathrm{rebr}$ after $t_0$. Note that SN~2022esa did
not show such a feature, so its ubiquity is unclear; we discuss
the effect of the rebrightening feature in Sec.~\ref{sec:parameterSpace}.

In principle, the amplitudes of the SN peak,
the oscillations, and the rebrightening feature are all related to 
the initial explosion and the distance. However, we have
decoupled the parameters to minimize assumptions about
the physical conditions. 

\begin{figure*}[ht]
    \centering
    \includegraphics[width=\textwidth]{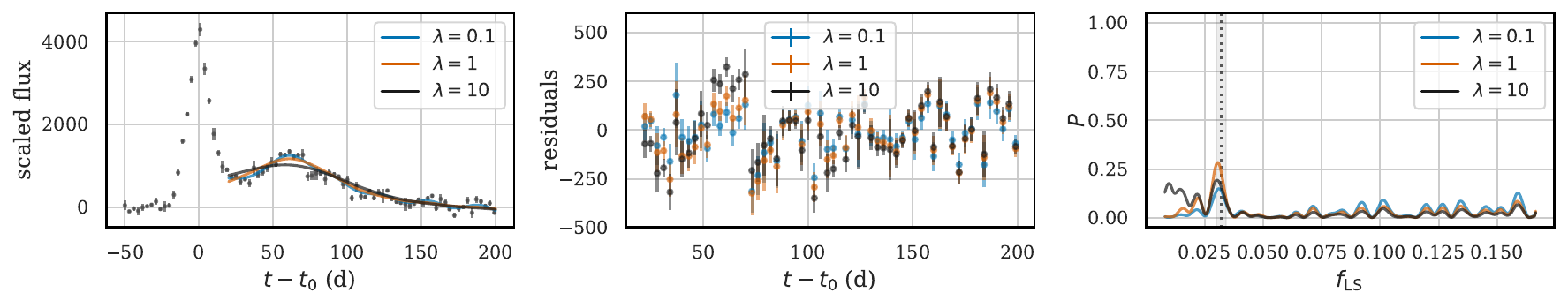}
    \caption{Lightcurves and spline fits (left), residuals (middle),
    and periodograms (right) for three values of the
    spline fit $\lambda$ on an example simulated SN with rebrightening
    and periodic features (see Sec.~\ref{sec:splines}). 
    The oscillations have a period of $T_\mathrm{osc} = 31.04$~d,
    which is shown as the vertical dotted line in the right panel, as well as a 
    shaded region corresponding to $f_\mathrm{osc} \pm 0.0021$ (the recovery
    condition). $\lambda$ controls the spline
    stiffness, and is an important parameter in our search. The loosest
    spline fit ($\lambda = 0.1$, blue) tends to subtract out
    the oscillations and reduces the periodogram power around $f_\mathrm{osc}$,
    while the stiffest spline fit ($\lambda = 10$, black) retains more of
    the underlying SN lightcurve itself, so that there is excess periodogram
    power at smaller frequencies / longer periods.
    Note that this is one
    particular example and conclusions on the optimal $\lambda$ cannot necessarily
    be generalized to the entire
    parameter space.}
    \label{fig:examples_lambda}
\end{figure*}

\section{Search procedure development}
\label{sec:development}

After simulating the SN population described in Sec.~\ref{sec:simulations}, 
we use spline fits to subtract out the underlying supernova
and isolate the periodicities, then apply a periodogram to the residuals. 
As our goal is to set up a realtime search using the AMPEL framework \citep{AMPEL}, we 
chose tools that are either already implemented or easy to implement
within this framework. 
With this in mind, we use the \texttt{scipy} \citep{scipy} functions
\texttt{make\_smoothing\_spline()}\footnote{\url{https://docs.scipy.org/doc/scipy/reference/generated/scipy.interpolate.make_smoothing_spline.html}} 
and its implementation
of the generalized Lomb-Scargle periodogram \footnote{\url{https://docs.scipy.org/doc/scipy-1.16.2/reference/generated/scipy.signal.lombscargle.html}}.
This
function takes a set of angular frequencies $\{2\pi f_\mathrm{LS}\}$ as the input frequency grid.

A three-day cadence was standard for the ZTF
public Northern sky survey in Phase I (before September 2020), 
with the equivalent Phase II survey having a 
two-day cadence\footnote{\texttt{https://www.ztf.caltech.edu/ztf-public-releases.html}}. 
In this section, we focus on
a three-day cadence and discuss the effects of a
two-day cadence in Sec.~\ref{sec:cadence}.

In contrast to most studies involving periodograms, which
are focused on single objects with long observations,
we are instead interested in a large population of objects with
limited observations. Our approach to the
problem as well as our search design choices 
keep this in mind.

\subsection{Periodogram setup}
\label{sec:grid}

When discussing the periodogram, we will refer to both the
search period $T_\mathrm{LS}$ and the the frequency $f_\mathrm{LS} = 1/T_\mathrm{LS}$.
$f_\mathrm{LS}$ is the more fundamental quantity,
while
$T_\mathrm{LS}$ lends itself to
more intuitive interpretations
as the period can be more easily extracted from visual examination
of a lightcurve.

For the periodogram, the most important input 
is the frequency grid, defined
by a minimum frequency $f_\mathrm{LS}^\mathrm{min}$, maximum frequency
$f_\mathrm{LS}^\mathrm{max}$, and frequency spacing $\Delta f_\mathrm{LS}$.
Equivalently, it is defined by $T_\mathrm{LS}^\mathrm{min}$ and
$T_\mathrm{LS}^\mathrm{max}$ over a grid that is more finely
spaced near $T_\mathrm{LS}^\mathrm{min}$ than $T_\mathrm{LS}^\mathrm{max}$.
$T_\mathrm{LS}^\mathrm{min}$ is usually recommended to be twice the
cadence to avoid undersampling; 
however, since the cadence is not completely regular, 
the strict Nyquist limit does not 
apply (e.g., Sec.~4 of \citet{LombScargle_VanderPlas}). 

The presence of two frequencies --- the signal frequency $f_\mathrm{osc}$
and the observing frequency $f_\mathrm{obs}$, or the inverse of the cadence --
results in aliases at beat frequencies of $f_\mathrm{osc}$ and $f_\mathrm{obs}$, as well as 
their higher order harmonics. These aliases occur at
\begin{align}\label{eq:aliases_f}
    f_\mathrm{alias} = \left| m f_\mathrm{osc} - n f_\mathrm{obs} \right|
\end{align}
where $m > 0$ and $n$ are integers. This effect is illustrated in Fig.~\ref{fig:example_aliases}.
For our studies, we found it sufficient to only consider $m=1$. The aliases
also give us a way to recover signals with
periods outside of our grid: Signals with periods less than $T^\mathrm{min}_\mathrm{LS}$
can be recovered at aliased frequencies (e.g., Eq.~47 in \citet{LombScargle_VanderPlas}\footnote{Note that we have
switched the sign compared to the notation in 
\citet{LombScargle_VanderPlas} in order to make the connection 
to beat frequencies more explicit.})
\begin{align}\label{eq:aliases_T1}
    T_\mathrm{recov} = \left| \frac{1}{T_\mathrm{osc}} - n f_\mathrm{obs} \right|^{-1}
\end{align}
with $n > 0 $, which comes directly from 
Eq.~\ref{eq:aliases_f}. 
We found $n = 1,2$ relevant for our search, which is related
to the periodogram grid frequency range versus
the range of simulated frequencies. 
Fig.~\ref{fig:example_lam1spl20_2Dhist} shows an example of
aliased signals in frequency space. 

For our search, we set $T_\mathrm{LS}^\mathrm{min}$ to be six days.
$T_\mathrm{LS}^\mathrm{max}$ should
be chosen to be both theoretically relevant and detectable
within the limits of our data sets, so we choose a maximum of 125
days. As recommended by \citet{LombScargle_VanderPlas}, we use
a frequency grid spacing of $\Delta f_\mathrm{LS} = 1/n_0 t_\mathrm{obs}$
where $t_\mathrm{obs}$ is
the observation duration and $n_0$ --- which determines how
well sampled a given peak is --- is commonly set to 
five or ten \citep{LS_n0_1, LS_n0_2, LS_n0_3, LS_n0_4}. 
With $n_0 = 10$ and $t_\mathrm{obs} = 200$, we have a grid spacing of
$\Delta f_\mathrm{LS} = 0.0005$~d$^{-1}$, for a total of 318
gridpoints per SN. Given
a fixed search grid, searches on data with $t_\mathrm{obs} < 200$~d 
are equivalent to larger values of $n_0$ and therefore a larger degree
of oversampling, without affecting the statistical meaning
of any result.

We apply the least-squares normalization to
the periodogram (the \texttt{normalize=True} option),
so that $P$ is in $[0,1]$ with higher values indicating larger
signal-to-noise ratios.
In principle, this allows for a straightforward comparison
of all periodogram peaks, with the detection criterion being some
threshold on the height of the tallest periodogram peak $P_\mathrm{max}$.
$f_\mathrm{recov}$ would then be the frequency $f_\mathrm{LS}$ 
associated with $P_\mathrm{max}$. However, when the number of data 
points $N_\mathrm{obs}$ is small, 
spurious peaks from noise 
fluctuations will be at similar heights to a real signal peak
(e.g., Fig.~25 in \citet{LombScargle_VanderPlas}). 
This means that a threshold
on (for instance) the highest peak needs to take $N_\mathrm{obs}$
into account. The effect is explored in Sec.~\ref{sec:durations}.

In principle, periodogram grids with smaller 
$T_\mathrm{LS}^\mathrm{min}$ could potentially improve
the sensitivity to shorter period signals, with the downside
of higher computational cost. In this scenario,
some longer period signals would instead be recovered at shorter
period aliases. We also tested grids with $T_\mathrm{LS}^\mathrm{min} = 3$~d 
and 1~d, but found these wider grids did not yield an improvement
in search sensitivity. This is further discussed in Sec.~\ref{sec:shorter}.

\subsection{Spline fits}
\label{sec:splines}

\begin{figure}[ht!]
    \centering
    \includegraphics[width=0.95\hsize]{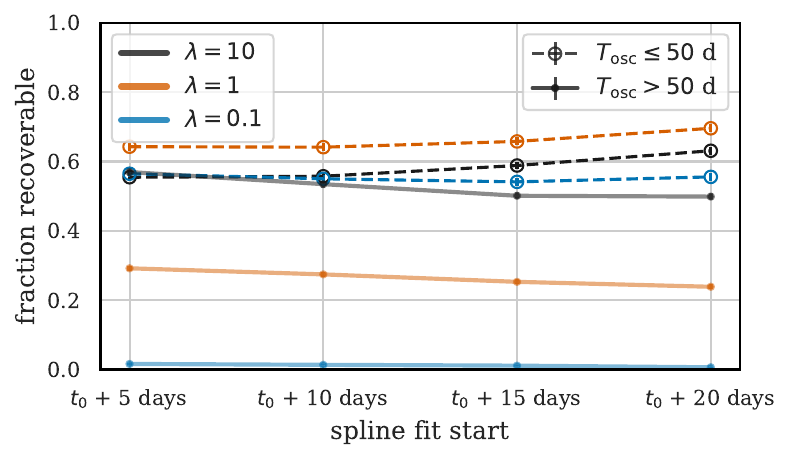}
    \caption{Fraction of recoverable signals using spline fits with different
    values of $\lambda$ and start times on the set of 10 000 simulated SN
    lightcurves. These are divided into the set of shorter period oscillations
    (open circles) and longer period oscillations (closed circles). 
    The relative behaviors of the curves are more meaningful than the 
    absolute fractions. For the stiff splines ($\lambda = 10$, black), 
    the choice of the spline fit start makes a large difference
    in signal recoverability, while for the loose splines ($\lambda =0.1$, blue)
    the difference is minimal. The moderate spline ($\lambda = 1$, red) has the 
    best performance for oscillations with $T_\mathrm{osc} \leq 50$~d while
    the stiff spline has the best performance for the slower oscillations.}
    \label{fig:testingSplineFits}
\end{figure}

In order to subtract out
the underlying lightcurve and extract
the oscillations, we need a function that is agnostic
as to the specific shape 
that would not subtract
out the periodicities themselves. Spline fits match
these requirements: they include a tuning
parameter on how rigid or flexible they are 
and are computationally inexpensive, making them
a good choice for a realtime implementation. We used
the \texttt{make\_smoothing\_spline()} implementation in \texttt{scipy},
which takes a penalty factor $\lambda \ge 0$,
with larger (smaller) values of $\lambda$ corresponding to stiffer
(more flexible) splines. We use $1/\sigma_F^2$ for the 
weights where $\sigma_F$ is the uncertainty in the flux. 
The residuals after subtracting
the spline fit from the lightcurve are then
processed through the Lomb-Scargle periodogram.

To help guide intution in the choice of $\lambda$, in Fig.~\ref{fig:examples_lambda} 
we show examples
of $\lambda = [0.1, 1, 10]$ on the same simulated $g$-band 
lightcurve. For the simulation, we used the \texttt{snana-2004gv} template
(a smoothly decaying type Ib template without much short-term
variability) and a three-day cadence. 
We set the oscillation
period to be $T_\mathrm{osc} = 31.04$~d with a scaled amplitude of 101.3.
We included all data between 20~d and 
200~d in the spline fits. As can be seen in Fig.~\ref{fig:examples_lambda},
the $\lambda=1$ spline fit results in the largest $P_\mathrm{max}$.
The $\lambda=10$ spline fit leaves additional structure in the residuals,
%
while the $\lambda=0.1$ fit
partially subtracts out the amplitudes themselves. While
this is a single example and does not necessarily generalize
to the entire population, it illustrates the differences
resulting from the different values of $\lambda$.
The choice of an appropriate
$\lambda$ is a well-studied problem, although common
prescriptions for finding an optimal $\lambda$ are not directly
applicable here since we are not aiming for a spline that
fully predicts the data. A more rigorous optimization of
$\lambda$ is outside the scope of this paper; instead, we tested
a few different values on our simulated lightcurves 
to explore its effect. We also tested a few different
start times after $t_0$ for the spline fit.

For a given simulated SN's periodogram, we define the SN
as recoverable if the frequency $f_\mathrm{recov}$ associated
with the maximum periodogram peak $P_\mathrm{max}$
is within $\pm 0.0021$~d$^{-1}$ (i.e.,
4.2 grid points) of $f_\mathrm{osc}$. The choice
of 4.2 rather than an integer is to account for potential rounding
issues when converting from period to frequency. We also allow
for signals to be recovered if they are within $\pm 0.0021$~d$^{-1}$
of an alias of $f_\mathrm{osc}$, which we will refer to as the
case of aliased recovery.
Note that all of these signals should
be considered as potentially recoverable rather than 
simply recovered, as the latter
requires some criteria for recovery and a characterization of
false dismissal probabilities. Additionally,
the set of recoverable signals will include
false alarms produced by fluctuations.

\begin{figure}[ht!]
    \centering
    \includegraphics[width=0.9\hsize]{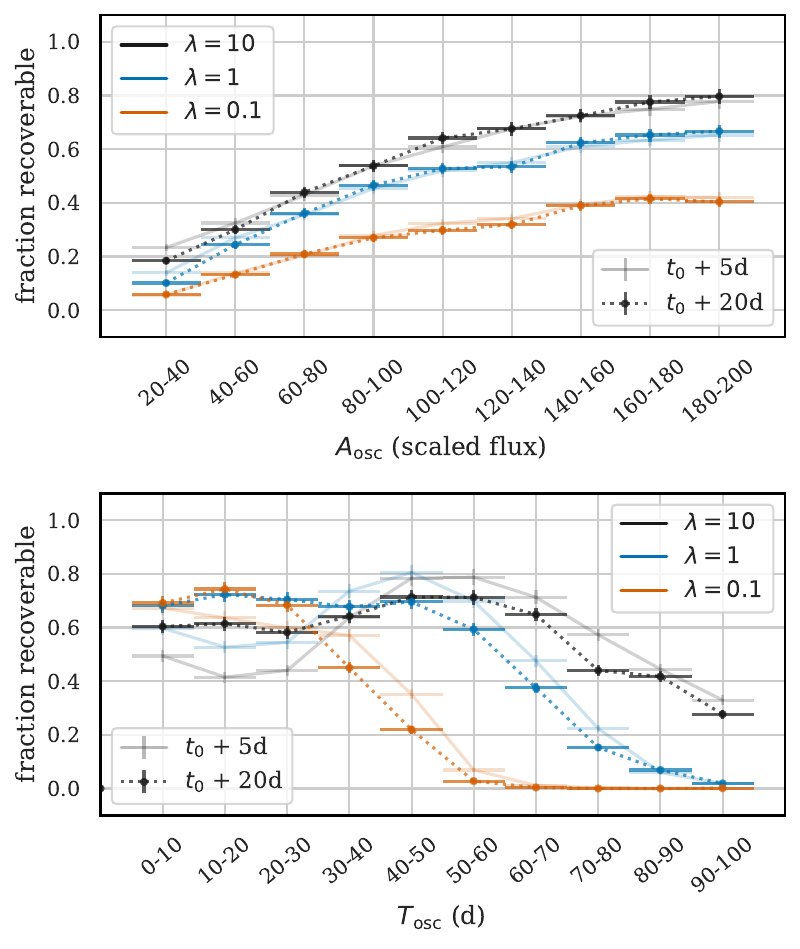}
    \caption{Fraction of recoverable signal using spline fits with different
    values of $\lambda$ and start times, as a function of oscillation amplitudes $A_\mathrm{osc}$
    (top) and periods $T_\mathrm{osc}$ (bottom) of the oscillations. For both figures, the
    more opaque symbols and lines correspond to earlier spline fit starts (5 days after $t_0$)
    while the more transparent symbols and lines correspond to later starts
    (20 days after $t_0$). Vertical error bars show the counting uncertainties. The shortest period oscillations ($T_\mathrm{osc} < 30$~d)
    are best recovered with the $\lambda = 0.1$ spline while the longest period
    oscillations ($T_\mathrm{osc} > 60$~d) are best recovered with
    $\lambda = 10$. The $\lambda = 1$ spline performs well across most
    of the range but loses sensitivity when $T_\mathrm{osc} > 60$~d.}
    \label{fig:testingSplineFitsDetails}
\end{figure}

We tested spline fits with combinations of $\lambda \in [0.1, 1, 10]$
and starting $[5, 10, 15, 20]$ days after $t_0$. Overall, 
the performances and recommendations are different for
shorter versus longer period oscillations
(Fig.~\ref{fig:testingSplineFits}). For signals with
$T_\mathrm{osc} < 50$~d, the moderately stiff spline
with $\lambda = 1$ produces the largest number of recoverable
signals, with a similar but slightly better performance
when starting later rather than earlier.
In contrast,
for $T_\mathrm{osc} > 50$~d, the stiff spline with
$\lambda=10$ greatly outperforms both the $\lambda = 1$
and $\lambda=0.1$ splines. The $\lambda = 0.1$ spline
fails to recover any longer period oscillations, as these
are generally all subtracted out by the loose spline fit.

For all three choices of $\lambda$,
the performance improves with increasing $A_\mathrm{osc}$ and
begins to level out at scaled fluxes $>\!\!160$
(Fig.~\ref{fig:testingSplineFitsDetails}, top). 
(For reference, this
corresponds to a magnitude difference $\Delta m = 0.04$ at 16 mag
and 0.25 at 18 mag.) 
Note that the exact percentages presented 
in this and the next section are not
particularly relevant, as we are free to choose the maximum
$A_\mathrm{osc}$ to scale the number of recoverable
signals. Instead, the relative behavior between the different
choices is the interesting property.

The recoverable percentages for the $r$ band were consistently worse
than for the $g$ band for $\lambda = 1, 10$ but slightly better for
$\lambda = 0.1$, with the differences being more pronounced for
signals with $T_\mathrm{osc} \leq 50$~d 
(Fig.~\ref{fig:splineFits_bands}). 
Since the rebrightening and oscillation
features are the same for both bands, this suggests that the supernova
templates themselves vary more rapidly in the $r$- than the $g$-band.

For the rest of this section, we will use the spline
fits with $\lambda=1$ starting at $t_0 + 20$ days. For comparison,
we will also run the $\lambda=10$ spline fit starting at $t_0 + 20$
days.

\subsection{Recoverable parameter space}
\label{sec:parameterSpace}

\begin{figure}
    \centering
    \includegraphics[width=0.9\hsize]{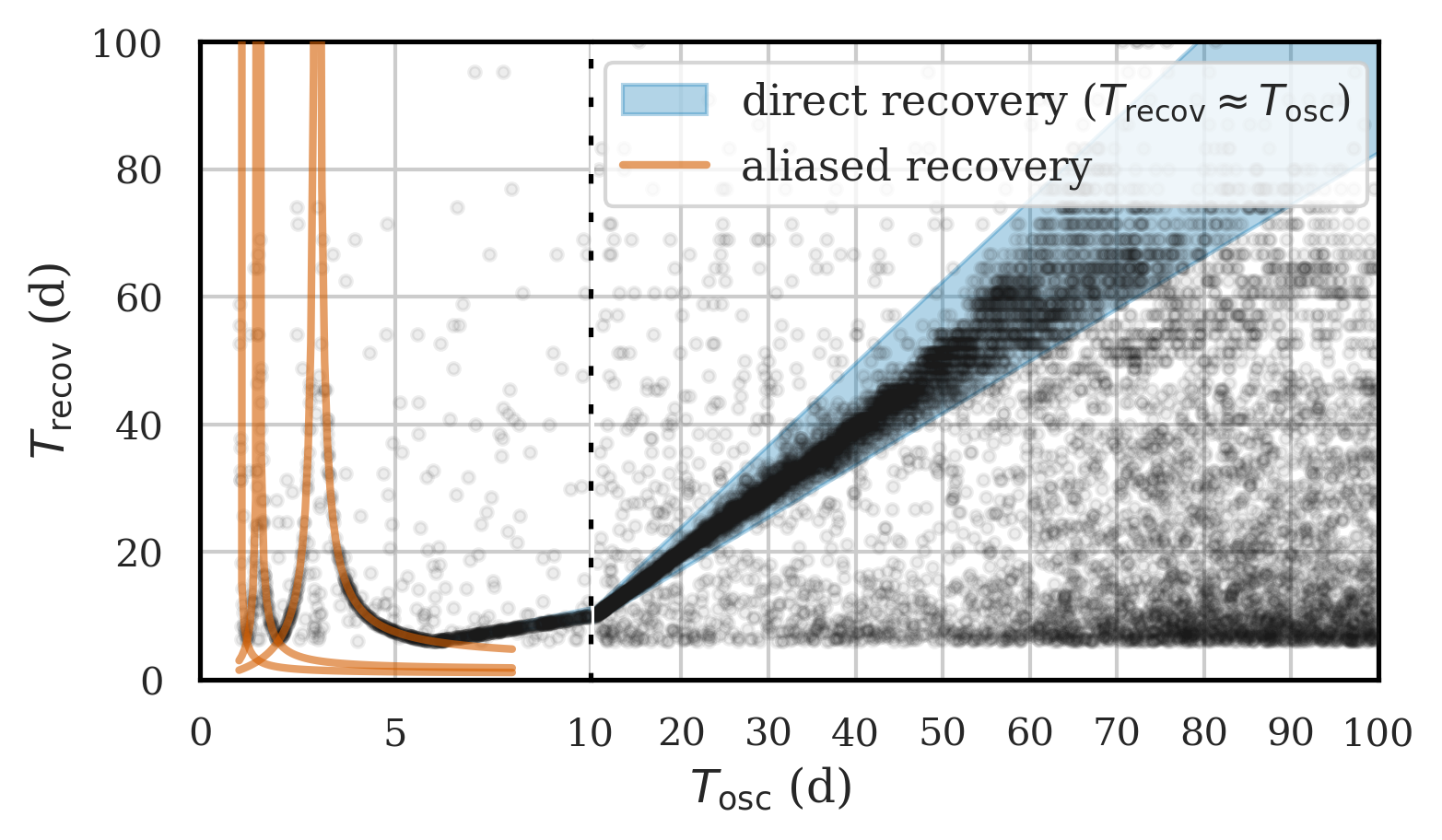}
    \caption{Recovered ($T_\mathrm{recov}$) versus true
    ($T_\mathrm{osc}$) periods in a search with 
    $T^\mathrm{min}_\mathrm{LS} = 6$~d, where the simulated SN
    lightcurves (gray circles)
    are observed with a three-day cadence. The $x$-axis for 
    $T_\mathrm{osc} < 10$~d
    is zoomed in to better illustrate the short-period signals that are recovered
    at longer aliased periods (Eq.~\ref{eq:aliases_T1}), which fall along
    the orange curves for $T_\mathrm{osc} < 6$~d. For $T_\mathrm{osc} > 6$~d, there is a 
    clear excess that corresponds to $f_\mathrm{recov} = f_\mathrm{osc} \pm 0.0021$
    (blue shaded region). The equivalent figure in frequency
    is shown in Fig.~\ref{fig:example_lam1spl20_2Dhist}.
    }
    \label{fig:scatter_6days}
\end{figure}

\begin{figure*}[ht!]
    \centering
    \includegraphics[width=\textwidth]{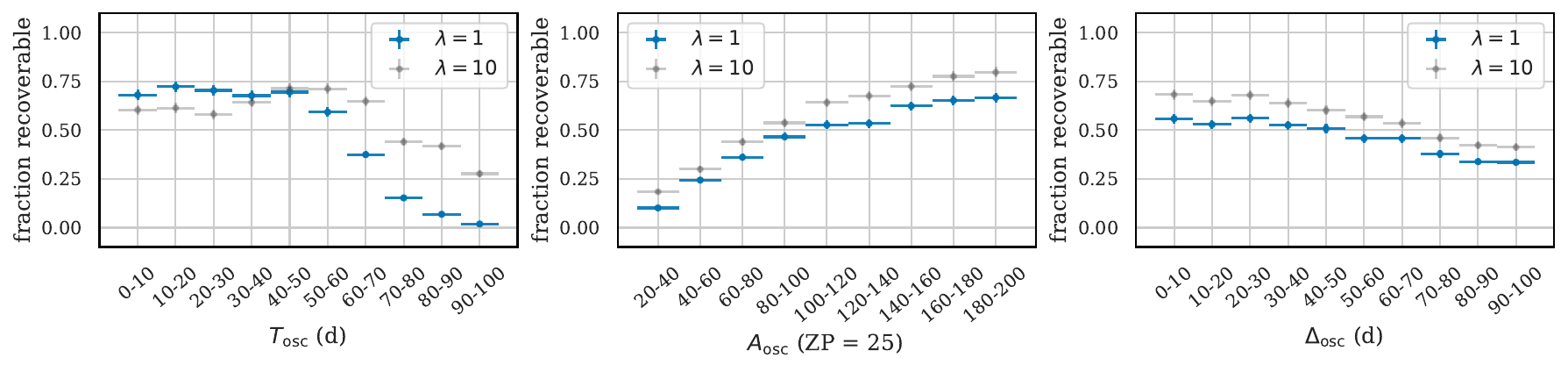}
    \caption{The fraction of recoverable signals as a function of the 
    oscillation parameters $T_\mathrm{osc}$ (left), $A_\mathrm{osc}$ (middle),
    and $\Delta_\mathrm{osc}$ (right) for the standard search with $\lambda = 1$
    (blue) and $\lambda = 10$ (black). 
    The $\lambda = 1$ search is less sensitive to 
    signals with longer $T_\mathrm{osc}$;
    both have more difficulty recovering signals with 
    smaller $A_\mathrm{osc}$ and larger $\Delta_\mathrm{osc}$.}
    \label{fig:triple_oscParams}
\end{figure*}

It is important to verify the parts of the parameter space that the
search method is most sensitive to. Figure~\ref{fig:scatter_6days} shows
the recovered $T_\mathrm{recov}$ versus the simulated $T_\mathrm{osc}$
for the 10 000 SNe, with the $\lambda = 1$ spline fit starting at $t_0 + 20$~d.
For $T_\mathrm{osc} > 6$~d, there is a
clear excess around $T_\mathrm{recov} = T_\mathrm{osc}$ with a spread
that corresponds to $f_\mathrm{recov} = f_\mathrm{osc} \pm 0.0021$~d$^{-1}$. 
A set of signals with $T_\mathrm{osc} < 6$~d are recovered
at aliased frequencies (orange curves);
these patterns can be more easily seen in frequency space 
(Fig.~\ref{fig:example_lam1spl20_2Dhist}).
The cluster of points around $T_\mathrm{osc} = 3$~d and low
$T_\mathrm{recov}$ show the difficulty in recovering 
signals with the same period as the search cadence. 

The search with $\lambda = 1$ has difficulty recovering
signals with $T_\mathrm{osc} > 50$~d (Fig.~\ref{fig:triple_oscParams}, left),
as $P_\mathrm{max}$ is instead found at $T_\mathrm{recov} < T_\mathrm{osc}$
(Fig.~\ref{fig:scatter_6days}).
In contrast, the search with $\lambda = 10$ performs comparatively better 
for these longer
period oscillations but worse for the shorter period ones. 

As expected, signals with larger amplitudes $A_\mathrm{osc}$ and shorter
delays $\Delta_\mathrm{osc}$ are more likely to be recovered. While we
limited our simulations to $A_\mathrm{osc} < 200$ and $\Delta_\mathrm{osc}
< 100$~d, it is clear that oscillation signals that are louder than this
limit are more likely to be recovered, while delays longer than this
are less likely to be recovered as fewer full oscillation
cycles are observed. 
Of course, a longer observation duration $t_\mathrm{obs}$ 
would allow for scenarios with longer delays and periods to be explored.

For the search with $\lambda = 1$, none of the rebrightening parameters
had an effect on signal recovery (Fig.~\ref{fig:tripleScatter_rebrParams}).
In fact, when we reran the search on a set of simulated lightcurves without
the rebrightening feature (i.e., SN and oscillations only), the fraction
of recoverable signals remains the same. 
For the $\lambda = 10$
search, the overall percentage of recoverable signals increased by
5 or 10\% when the rebrightening features were removed. 
This indicates that the $\lambda=1$ spline is flexible enough
to accommodate the rebrightening features while the $\lambda=10$ spline 
is moderately affected by them, and searches that use stiffer splines 
(e.g., to be more sensitive to longer period oscillations) will
have to account for this by, e.g., excluding more early data.

\section{Results}
\label{sec:results}

\subsection{Observational gaps}
\label{sec:gaps}
Approximately
10\% of scheduled ZTF observations are canceled
due to external reasons such as bad weather.
When the lost observations are randomly distributed, 
the effect of this is minimal (Fig.~\ref{fig:dropdays}): 
When we dropped a random 10\% of observations, the
the standard search with $\lambda = 1$ was unable to recover
only 1.1\% of the previously recoverable signals. 
(With a more pessimistic 30\%
drop rate, 4.3\% are no longer recoverable. For the search
with $\lambda = 10$, the reduction in the 
recoverable signals is 1.3\% and 6.1\% for the 10\% and 30\%
drop rates, respectively.) 
We will therefore apply a random 10\% loss of observations 
to the searches on simulated data in this section.

\begin{figure}[ht!]
    \centering
    \includegraphics[width=0.9\hsize]{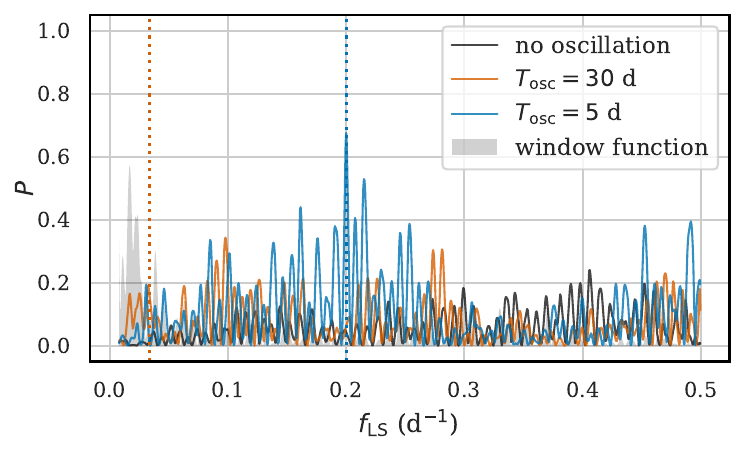}
    \caption{Periodograms for three lightcurves with the same 
    observation dates, taken from the $g$-band observations of
    SN~2018grf (see top row, right panels of Fig.~\ref{fig:ZTF_mid_examples}),
    which includes multiple long gaps. The power in the window function
    itself is shown by the shaded region.
    When there is no additional oscillatory signal (black curves),
    the periodogram shows no strong features. However, when an additional
    bright 5-day (blue) or 30-day (orange) oscillation is included, there
    is additional structure due to the interplay between the
    signal periods and the gap timescales. The vertical dotted lines mark
    $f_\mathrm{osc}$ for the two signals.}
    \label{fig:example_gaps_structure}
\end{figure}

In reality, the canceled observations
are more likely to appear consecutively due to external
conditions that span multiple days. These longer
gaps would decrease the sensitivity to
signals and increase the number of false alarms. In particular,
multiple long gaps can add an additional timescale to the system and
lead to more complex features in the periodograms. 
Fig.~\ref{fig:example_gaps_structure} illustrates this by comparing
the periodograms for three lightcurves with the same observation
dates, which are taken as the dates of the $g$-band observations
of SN~2018gf (to be discussed in Sec.~\ref{sec:ztf}). 
This is a SN
with multiple long gaps in its lightcurve: a 56-day gap, a 27-day gap, 
and multiple gaps of 5-15 days. When the lightcurve contains no additional
periodic signal (black curves), the periodogram does not show any
strong features. However, for the lightcurves with additional
periodic signals with $T_\mathrm{osc} = 5$~d (blue curves) or
$T_\mathrm{osc} = 30$~d (red curves), the periodograms show more
complex features and many high-power peaks beyond the ones expected
at $f_\mathrm{osc}$. 
The periodogram for the window function ---
calculated by applying the periodogram to a set of constant
values at the observation times of the SN --- only shows 
the peaks around 1/56 and 1/27, whereas the structure in the lightcurve
periodograms comes from the combination of the multiple timescales.

In this particular example, the 5-day period
signal is recovered but the 30-day period signal is not; however,
this is a specific example and does not necessarily generalize.
The same
will also occur when there is any short-duration variability, real
or instrumental. Searches on real data will therefore have
reduced sensitivity when structured gaps are present.

\subsection{Total observation durations}
\label{sec:durations}

So far, we have assumed a very optimistic 200
days of observations after $t_0$ for all SNe, but in reality,
the majority of ZTF SNe are observed for a month before the planned
schedule moves on to other parts of the sky, and only particularly
bright or otherwise interesting SNe are observed longer.
Shorter observation durations reduce the number of
recoverable signals in multiple ways: The sensitivity
reduces overall when fewer full cycles are observed, and the sensitivity
to longer period signals and signals with longer delays $\Delta_\mathrm{osc}$
is especially reduced. In addition, with shorter observation durations
--- or, more precisely, a small number of data points $N_\mathrm{obs}$ ---
the periodogram peaks produced by noise fluctuations 
are taller.

\begin{figure}[ht!]
    \centering
    \includegraphics[width=0.9\hsize]{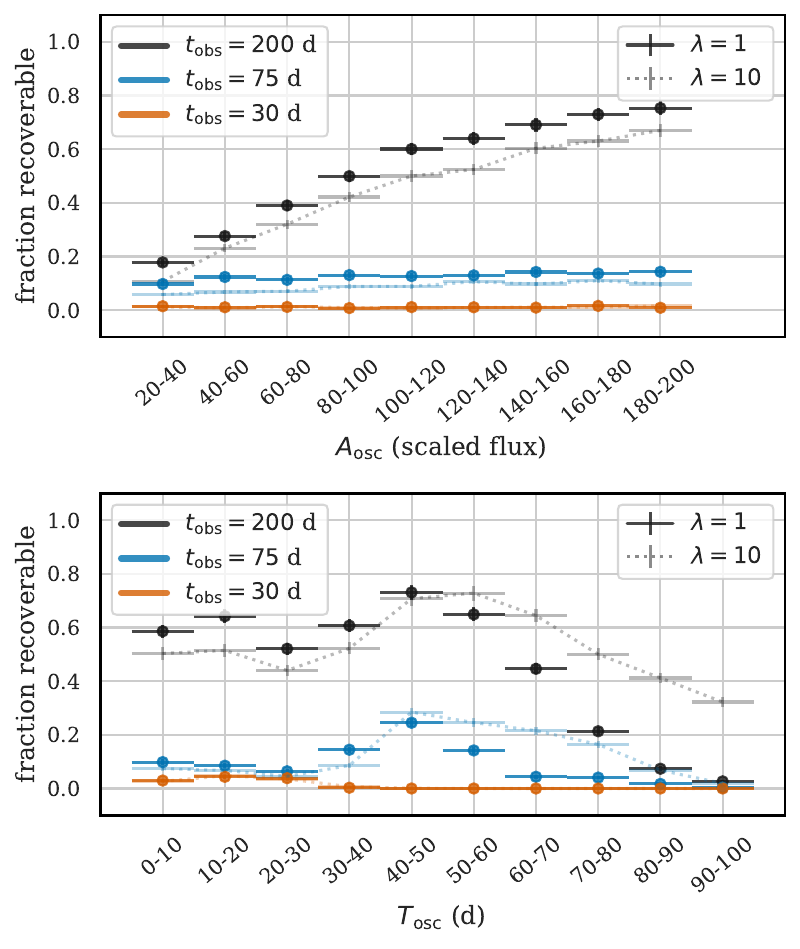}
    \caption{The fraction of recoverable signals drops dramatically
    when the total observation duration drops from 200 days
    (black) to 75 days (blue) and further
    still to 30 days (red), assuming the same three-day
    cadence. For a 30-d observation, only short-period signals
    are recoverable, and only a very small fraction of them. 
    For the 75~d observation, the recoverable signals with 
    $T_\mathrm{osc} > 40$~d are in fact almost all due to noise fluctuations. 
    For both of the shorter observation
    durations, the inclusion of brighter oscillations does not improve the
    recoverable fraction.}
    \label{fig:passed_durations}
\end{figure}

To quantify the reduced sensitivity with shortened observation durations, we
reran the search using $t_\mathrm{obs} = 30$~d and 75~d and compared them
to the search with $t_\mathrm{obs} = 200$~d. For all of the searches
in this section,
we started the spline fits at $t_0 + 10$~d rather than 20~d, to ensure that
we have enough data points for the spline fit.
If too many observations were dropped in the random 10\% loss
due to weather, successively earlier data points were included until
enough points remained.

Figure~\ref{fig:passed_durations} shows the fraction of recoverable signals
for the three $t_\mathrm{obs}$ values and the two $\lambda$ settings. 
For $t_\mathrm{obs}=200$~d, the signal recovery improves with brighter
signals; in contrast, the recovery rates for the other two remain relatively 
constant, as the bottleneck here is the observation duration. 
The $t_\mathrm{obs}=30$~d search is unable to recover any but the shortest
period signals with short delay times. 
For $t_\mathrm{obs}=75$~d, the increase in recoverable signals at $T_\mathrm{obs} > 40$~d is spurious:
the spline fit residual has a strong
initial excess from the tail of the SN peak, which results in periodogram
power at intermediate values of $T_\mathrm{recov}$.
This shows that simply considering whether
signals are recoverable can lead to misleading conclusions, and a discussion of
false positives is necessary.

\begin{figure}[ht!]
    \centering
    \includegraphics[width=0.9\hsize]{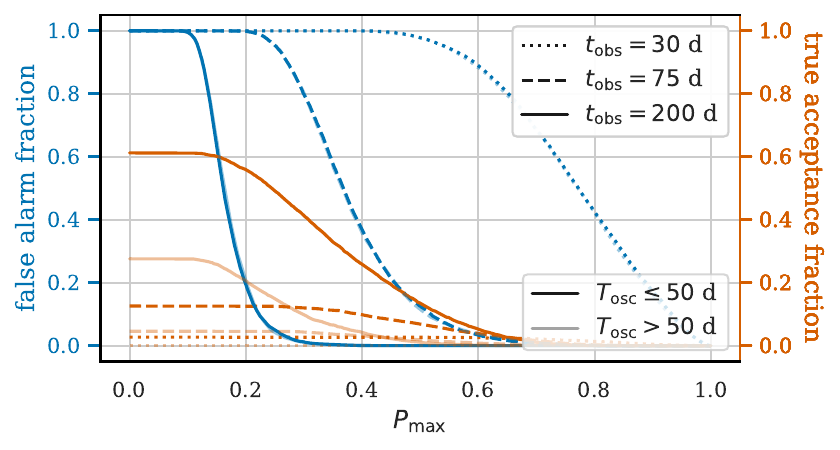}
    \caption{False alarm (blue) and true acceptance (orange) fractions
    for 30-, 75-, and 200-day observations with
    $\lambda=1$. The opaque curves
    correspond to the set of signals with
    $T_\mathrm{osc} \leq 50$~d
    and the more transparent curves to $T_\mathrm{osc} > 50$~d. (The false alarm
    curves are naturally the same for both subsets.) With longer durations,
    false alarm curves shift to the left while true acceptance curves
    shift upward.}
    \label{fig:FATA_threeDurations}
\end{figure}

\begin{figure}[ht!]
    \centering
    \includegraphics[width=0.85\hsize]{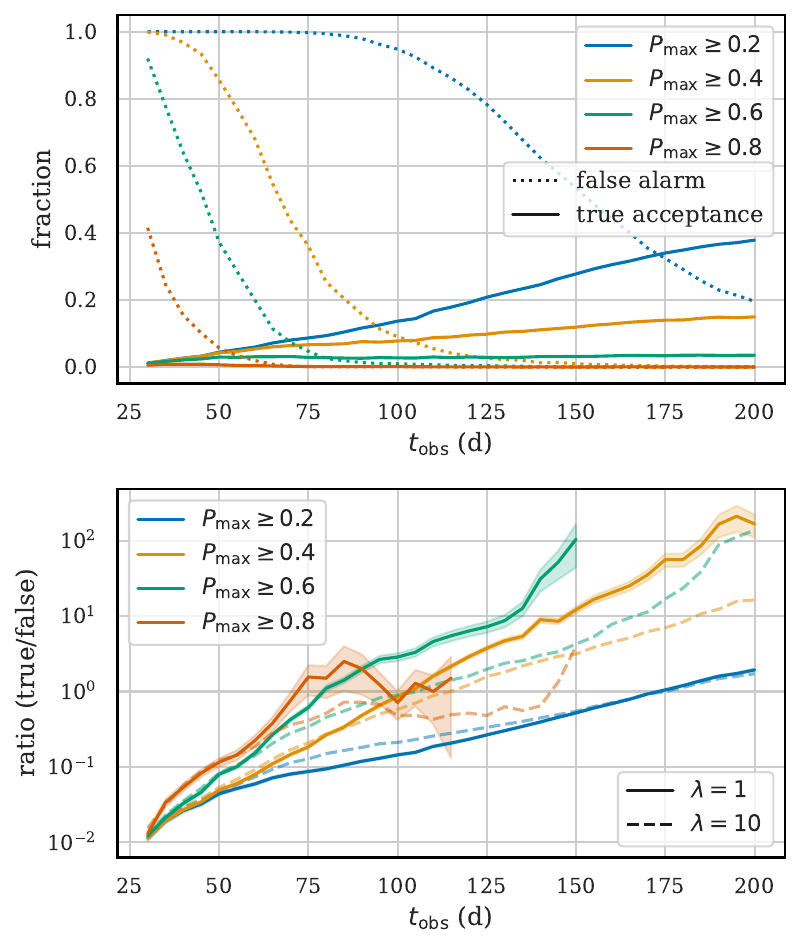}
    \caption{The false alarm and true acceptance (top)
    as well as the ratio (bottom) as a function of
    observation duration for four $P_\mathrm{max}$ thresholds
    with the $\lambda=1$ search. While
    the ratio improves for higher $P_\mathrm{max}$ thresholds, 
    a too-stringent threshold also precludes most of the signal.
    In the bottom plot, the shaded regions show the propagated
    uncertainty in the number counts, and the curves end where
    there are no more false alarms. In addition, the curves for
    the $\lambda=10$ search are plotted for easier comparison.}
    \label{fig:FATA_fourThresholds}
\end{figure}

Figure~\ref{fig:FATA_threeDurations} 
shows the false alarm and the true acceptance fractions for 
the $\lambda=1$ search for the three 
valeus of $t_\mathrm{obs}$. The false alarms are 
calculated by running the
same search on the SN-only lightcurves (i.e., without the oscillations
and rebrightening features) to determine $P_\mathrm{max}$.
In both figures, the true acceptance fraction is plotted
separately for signals with $T_\mathrm{osc} \leq 50$~d (opaque curves)
and $T_\mathrm{osc} > 50$~d (semitransparent curves).
The $\lambda=10$ search produces more false positives from the
residual excesses due to poor spline fits.
It is clear that a 30-day observation is insufficient to
recover almost any of the simulated parameter space, and
a 75-day observation still suffers greatly when compared to a 
200-day observation.

A threshold on $P_\mathrm{max}$ is necessary to define a detection.
Figure~\ref{fig:FATA_fourThresholds} (top)
shows how the false alarm and true acceptance fractions change
with $t_\mathrm{obs}$ for four different thresholds
on $P_\mathrm{max}$, for the $\lambda=1$ search. 
In general, the false alarm rates
decrease toward zero after a certain duration while the
true acceptance rates increase monotonically. While the ratio (bottom)
is generally better for higher $P_\mathrm{max}$ thresholds, the
number of signals also decreases, and an intermediate value
such as $P_\mathrm{max} \geq 0.4$ is a good compromise for
most search scenarios. 
When considering the $\lambda=10$ search, the ratio of
true signals over false alarms is generally lower due to the higher
prevalence of false positives when using a stiffer spline.

For any threshold above $P_\mathrm{max} \geq 0.2$, the number of false
alarms tends to become manageable at $t_\mathrm{obs} = 100$~d, as
this is where the curves start to flatten. This
suggests that, assuming a three-day cadence, at least 100~days of observations
are necessary for noise rejection. 
Longer observations will additionally increase the
chances of detecting a signal, particularly ones with longer periods
or delays. 

As the number of false alarms is mostly a function of 
$N_\mathrm{obs}$ on its own, this means that the false alarms
for a two-day cadence, for instance, will be suppressed at around
65~days of observations. In contrast, the number and parameter space
of recoverable signals also depends on the cadence; this is explored
further in Sec.~\ref{sec:cadence}.

\subsection{Follow-up observations}

\begin{figure}[ht!]
    \centering
    \includegraphics[width=0.9\hsize]{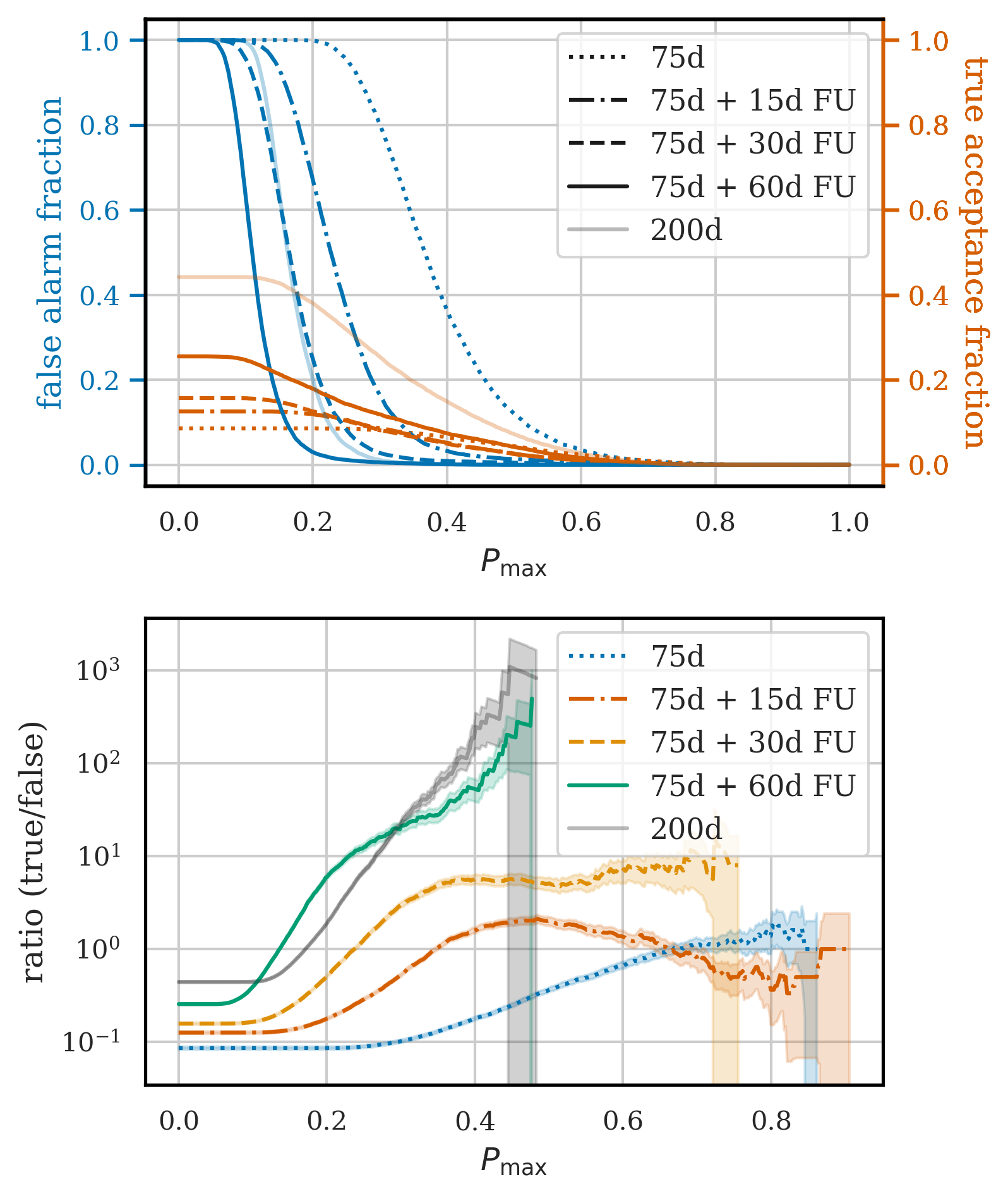}
    \caption{(Top) Similar to Fig.~\ref{fig:FATA_threeDurations} but comparing
    a few different follow-up (``FU'') scenarios for the $\lambda=1$ search.
    Starting from a base 75-day
    observation, we add 15, 30, and 60 days of follow-up observations
    at a one-day cadence. We also plot the curves from a full 200-day 
    observation for comparison. With 30 days of follow-up observations,
    the false alarm fraction is already similar to that for a
    200-day observation, although the number of true signals is still
    significantly lower.
    (Bottom) The ratio of true signals to
    false alarms is shown for the five different scenarios in the
    top panel. Adding follow-up observations greatly reduces the false
    alarm rate and therefore improves the ratio.}
    \label{fig:FATA_FUs_1d}
\end{figure}

While the typical SN is observed for less than
the recommended 100~days, higher cadence
follow-up observations can improve the performance.
To explore this, we start with 75 days of a three-day
cadence observation followed by three different follow-up durations
with a one-day cadence: a) 15 days, b) 30 days, and c) 60 days. 
From Fig.~\ref{fig:FATA_FUs_1d}, for the $\lambda=1$ search, it can 
be seen that 
the noise rejection (blue curves) with 30 days of follow-up
observations is as good
as for the 200-day search, despite having a smaller number of
observations ($N_\mathrm{obs} = 52$ versus 63). However, the sensitivity
to longer period signals and signals with longer delays will still be low
due to the shorter total observation times. The results are qualitatively
similar when considering the $\lambda=10$ search.

\subsection{Effect of the survey cadence}\label{sec:cadence}

\begin{figure}
    \centering
    \includegraphics[width=\hsize]{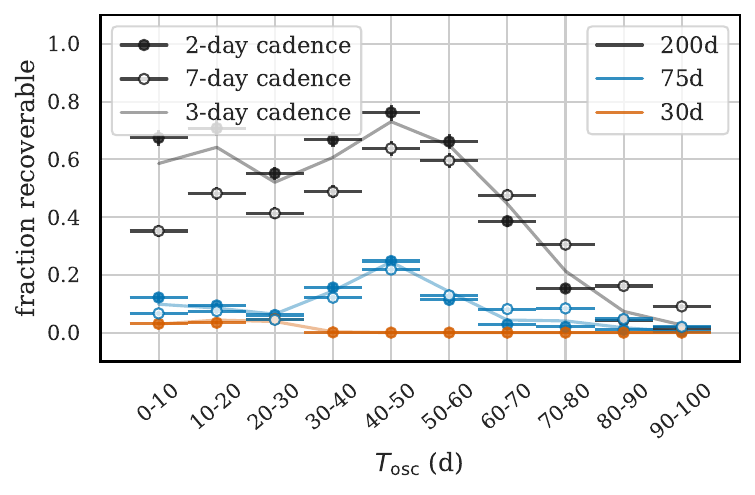}
    \caption{The recoverable fraction as a function of $T_\mathrm{osc}$
    for the two- (closed circles) and seven-day (open circles) cadence observations, compared
    to the three-day cadence (gray line). The two-day cadence performs overall better
    than the three-day cadence except for the longest period oscillations,
    while the seven-day cadence generally performs worse
    particularly for the shortest
    oscillations. As in Fig.~\ref{fig:passed_durations}, the recoverable
    fraction for the 75-day search at $T_\mathrm{obs} > 30$~d is spurious.}
    \label{fig:passed_cadences} 
\end{figure}

While we have so far focused on the three-day cadence
of the ZTF Phase I Northern Sky survey, 
we can also check how the results would change with
a faster (two-day) and slower (seven-day) cadence, particularly
the sensitivity to different values of $T_\mathrm{osc}$. A two-day
was used in the ZTF Phase II survey, while a seven-day
cadence is more representative of Legacy Survey of Space and 
Time (LSST) performed by the Vera C. Rubin Observatory \citep{LSST}.
Note that the choice of cadence affects the
minimum LS search period $T_\mathrm{LS}^\mathrm{min}$
as well as the definition of the aliased frequencies 
(Eq.~\ref{eq:aliases_T1}), which depend on the inverse of the cadence
$f_\mathrm{obs}$. For all the searches in this section, we set
$T_\mathrm{LS}^\mathrm{min}$ to be twice the cadence,
and maintained the grid spacing $\Delta f = 0.0005$~d$^{-1}$. We used
both $\lambda = 1$ and $\lambda=10$ and began the spline fit at $t_0 + 10$~days,

The sensitivity to signals generally increases slightly for a 
two-day compared to a three-day cadence
(Fig.~\ref{fig:passed_cadences}). 
However, the recovered fraction of signals with $T_\mathrm{osc} 
> 55$~d is slightly lower for a two-day cadence, possibly because
the longest period oscillations are more likely to be 
subtracted out by the spline fits when the cadence is higher.
This suggests that a search on higher cadence data should
consider stiffer splines (larger values of $\lambda$)
especially if the signals are expected to have longer
periods.

For the seven-day cadence, we did not include
any 30-day observations as this combination would not
provide enough data points for the spline fit.
While the number of
recovered signals is lower compared to the three-day search, 
the
sensitivity to short period signals with $T_\mathrm{osc} < 10$~d
is particularly bad. Interestingly, the recovery of the longest
duration signals is slightly better than the three-day
cadence search despite the lower number of observations.

The number of observations has
a large effect on the false alarm fraction
(Fig.~\ref{fig:FATA_cadences}). As expected, the
false alarms are suppressed with a two-day cadence and 
significantly more prevalent with a seven-day cadence.
These suggest that any conclusions about the false alarm
rate that have been drawn so far are specific to
a three-day cadence.

If the survey cadence is instead highly irregular, 
the signals will no longer be
recoverable at aliased frequencies. This suggests that irregular
surveys would need to use search grids that cover
a wider range of $T_\mathrm{LS}$ (in particular, smaller
values of $T_\mathrm{LS}^\mathrm{min}$) to ensure that
they contain the entire potential range of $T_\mathrm{osc}$,
which would increase computational cost. The effects of
a more irregular cadence in real data is shown
in the next section.

\section{Running the search on observed SNe}
\label{sec:realdata}

\subsection{Archival ZTF data}
\label{sec:ztf}

\begin{figure*}
    \centering
    \includegraphics[width=0.9\textwidth]{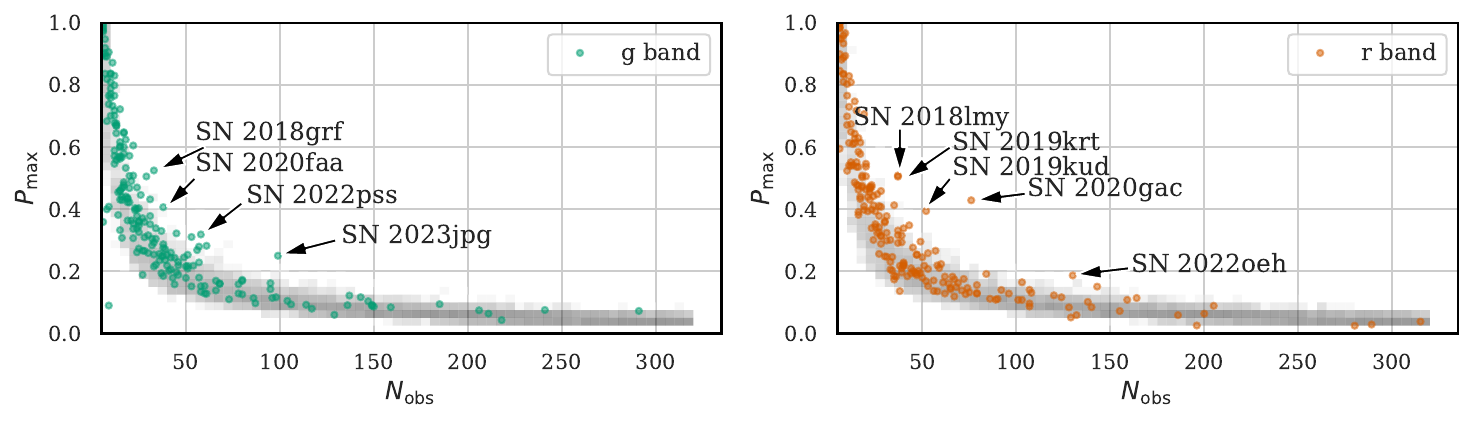}
        \caption{$P_\mathrm{max}$ vs. the number of data points 
        included in the spline fit $N_\mathrm{obs}$
        for the $g$ and $r$ bands individually, with the results
        from 10~000 simulations shown in the gray 2D histogram in the
        background (logarithmic scaling). The outlier SNe are
        labeled.
       The larger number
        of $r$-band outliers is likely due to the SNe being on average brighter
        and more variable in the $r$ than in the $g$ band. The lightcurves
        and periodograms for the labeled SNe, as well as a few 
        other examples, are plotted in Fig.~\ref{fig:ZTF_mid_examples}}.
    \label{fig:ZTF_mid_Pmaxs}
\end{figure*}

ZTF has surveyed the northern sky since 
2018 in multiple bands.
The ZTF Bright Transient Survey 
(BTS) \citep{ZTFBTS_Perley+}
aims to spectroscopically classify the subset of transients 
brighter than $\sim$18.7~mag. This dataset thus provides a 
large sample on which to test our search. We base 
our initial test on all CC SNe classified by BTS and fulfilling $0.03<z<0.06$, and with 
a fade timescale greater than $40$ days (as estimated by the BTS sample explorer), 
for a total of $212$ SNe as of late February 2026. The redshift range is 
set to match a realistic search volume while avoiding rare very nearby 
objects, and the fade timescale limit set to guarantee a minimal set of 
observations. 
For this sample we obtained 
IPAC Forced Photometry \citep{IPAC1, IPAC2}, which we processed 
based on IPAC 
recommendations (all observations flagged as cloudy were removed and 
observations in the secondary field grid were not used). Finally, 
extreme lightcurve outliers were removed using a sigma clip of $5\sigma$ with a
rolling window of 11 visits, using \texttt{rolling\_window\_sigma\_clip()} in the 
\texttt{fundamentals} package \citep{Young_fundamentals}. For a few
SNe, we manually filtered out some additional lightcurve outliers
that were not automatically removed in the previous step.

Most of the SNe in our dataset had average cadences of 
two or three days (Fig.~\ref{fig:ZTF_mid_cadences}), although
both longer and shorter delays between successive
observations are also present. Indeed, 
any individual
field could have a longer gap in coverage due to external
(e.g., weather) or scheduling reasons. 
In addition, our set of SNe also includes data from private programs, 
which often had higher cadences. 
Therefore, in reality the observations are better thought of as
having a base cadence of one day with some longer gaps, 
so we set $T^\mathrm{min}_\mathrm{LS}=2$~d. For lightcurves with
slower cadences, this means that the periodograms will contain
reflected portions (as illustrated in Fig.~\ref{fig:example_aliases}) and
any signals present could be at aliased frequencies.

We used the spline fit with $\lambda=1$ starting 20~days after
the peak (which is generally defined as the brightest $r$-band point)
and included all visits up to 200~days post-peak. As discussed, the 
number of points included in the periodogram
$N_\mathrm{obs}$ strongly affects $P_\mathrm{max}$. To have a comparison, we
also ran the same search on 10 000 new SN-only simulations
in the $g$ and $r$ bands separately with a one-day cadence
and dropping a random 10\% of observations. 
We modified the $z$ range to be
the same as the ZTF SN sample ($[0.03, 0.06]$) and set the
scaling parameter be in $[10^{-16}, 5\times10^{-15}]$
to produce SNe with peak fluxes
matching those of the ZTF sample. 

Out of these 212 SNe,
we identified nine with combinations of ($N_\mathrm{obs}, P_\mathrm{max}$)
that were outside of our simulated distribution  (Fig.~\ref{fig:ZTF_mid_Pmaxs}), 
which we highlight
as periodogram outliers. Although the choice of cutoff for identifying
an outlier is somewhat
arbitrary, these nine serve to illustrate the different
effects that might be seen in a real search. We did, however,
additionally inspect all the other periodograms manually.

\begin{figure*}[ht!]
    \centering
    \includegraphics[width=\textwidth]{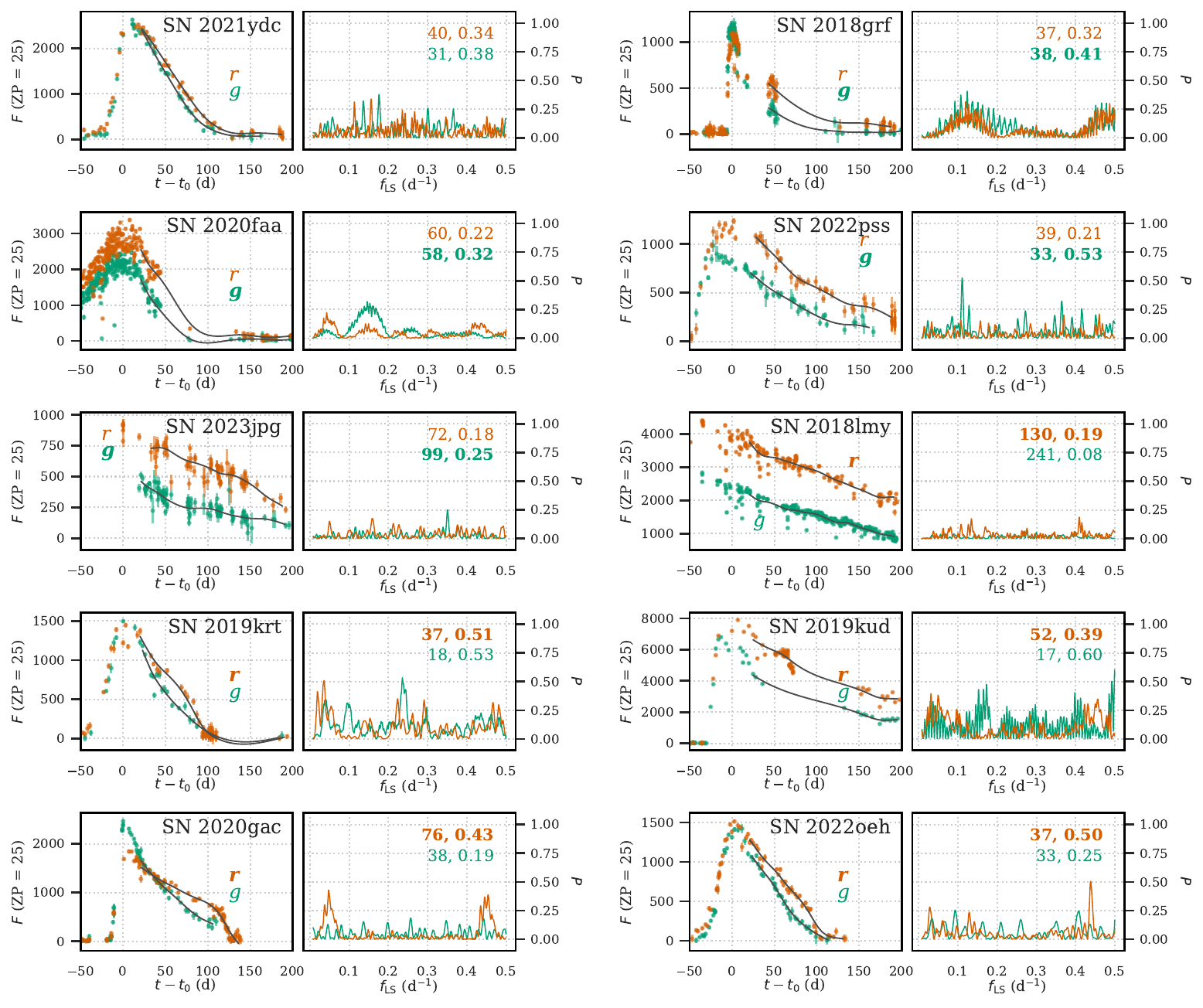}
    \caption{Lightcurves and spline fits (left) and periodograms (right) for
    ten SNe in the moderately bright and nearby ZTF sample. In the top-left
    panels, SN~2021ydc is an example of a well observed SN lightcurve, while
    the other nine are the four $g$-band and five $r$-band periodogram
    outliers from Fig.~\ref{fig:ZTF_mid_Pmaxs}, with the bold text in each
    panel indicating the band for which the SN is a periodogram
    outlier. The numbers on each periodogram are the values of $N_\mathrm{obs}$
    and $P_\mathrm{max}$ for the two bands. The most plausible signal candidates are
    arguably SN~2022pss and SN~2022oeh. For a few of these SNe, the periodogram
    appears reflected (e.g., SN~2022gac); for a periodogram search grid with
    $T_\mathrm{LS} = 2$~d, this indicates regular observations at slower than
    a one-day cadence.}
    \label{fig:ZTF_mid_examples}
\end{figure*}

A few lightcurves, spline fits, and periodograms are
shown in Fig.~\ref{fig:ZTF_mid_examples} for interesting SNe. Of these, SN~2021ydc
(top row, left) is an example of a regularly observed lightcurve 
without any notable periodogram peaks, while the other
nine are the periodogram outliers. SN~2018grf
(top row, right) shows the effect of the longer observation gaps
discussed in Sec.~\ref{sec:gaps}.
This effect also dominates
the periodograms of SN~2020faa, SN~2019krt, and SN~2019kud.
SN~2020faa and SN~2018lmy were also observed
in multiple fields and their lightcurves contains disparate values of
flux at similar times.
SN~2020gac has a poor spline fit at late times due to rapid changes 
in the SN lightcurve which drives the elevated $P_\mathrm{max}$;
this could indicate the start of interesting behavior such as
periodic oscillations but further observations would be necessary
to confirm this. The lightcurve of SN~2023jpg contains a lot of scatter, 
and the $P_\mathrm{max}$ seems to mostly be driven by a single
lightcurve outlier at around 120~days. The two remaining, SN~2022pss
($T_\mathrm{LS} \approx 9$~d)
and SN~2022oeh ($T_\mathrm{LS} \approx 2$~d), would be promising candidates for follow-up or
deeper observations, although both only present outlier periodogram peaks in
a single band. Both of these were classified as Type
IIn \citep{SN2022pss,SN2022oeh}, indicating a hydrogen-rich 
spectrum with circumstellar
interactions, which could explain any variability.

We also ran the search with $\lambda=10$ on these ZTF SNe. The $\lambda=10$
search found the same outliers as the $\lambda=1$ search as well as a few
additional ones. Upon manual inspection, none of the additional periodogram outliers 
presented any new interesting features.

Our search
method can also be applied to the brightest SNe, although with 
potential limitations. To explore this, we ran a search on 237
ZTF-detected CC SNe at $z < 0.03$. These had peak magnitudes as bright as 14  ---
equivalent to scaled fluxes of a few times
$10^4$ (ZP = 25) --- and up to $N_\mathrm{obs} > 500$,
so that new SNe simulations were necessary.
For this set of simulations, 
we set $z \in [0.001, 0.03]$
and the scaling parameter $\in [5\times10^{-14}, 10^{-13}]$.
We ran the 
search with the same data preparation, spline fits, and 
periodogam search grid as for the moderately bright sample. 

There were again more $r$- than
$g$-band periodogram outliers in the nearby SN sample
(Fig.~\ref{fig:ZTF_bright_Pmaxs}). The only
$g$-band outlier (SN~2018hna) was in fact an outlier in both bands, 
and both the large $P_\mathrm{max}$ values and $N_\mathrm{obs}$
were due to the SN being located in multiple fields. 
The other periodogram outliers were generally due to the spline fit not being
able to accommodate the rapid flux variations that can be observed
in bright SN lightcurves. 

Instead of using simulated SNe as a comparison, we could use
some false alarm probability calculation, which has the benefit of 
allowing for a straightforward quantitative cut to determine the SNe
of interest. However, the estimation of a false alarm probability
in real data is nontrivial (see Sec. 7.4.2 of
\citet{LombScargle_VanderPlas}). In practice, analytical approximations
of the false alarm probability depend on the properties of the
data set, particularly the noise distribution; for instance, 
the false alarm fraction is different for the $\lambda = 1$ and 
$\lambda = 10$ spline fits. Computational methods
such as bootstrapping are a promising alternative but 
can be prohibitively expensive for a realtime application. 

The larger number of periodogram outliers in the $r$ band
is at least partially due to the longer gaps that tended to be present
in the $r$ band compared to the $g$ band. As Fig.~\ref{fig:ZTF_mid_cadences} shows,
both the mean and the standard deviation in the delay between 
successive visits tend to be larger in the $r$ band, indicating not only
a slower cadence but also larger deviations from a regular cadence. It is 
also likely that the $r$-band lightcurves have more short term
variability, as noted in Sec.~\ref{sec:apprecovery}.

\subsection{Notable SNe}

Using our method, we are able to recover the 12.4~day periodicity for 
SN~2022jli. 
We obtained the photometric data from the public
repository associated with \citet{SN2022jli_Chen+}\footnote{\url{https://github.com/AtomyChan/SN2022jli}}. 
Using the standard procedure with a spline fit starting at $t_0 + 20$~d
and $\lambda = 1$, we recover a $g$-band $P_\mathrm{max} = 0.110$ for
$N_\mathrm{obs} = 346$ when we only include data up to 250 days
(after which point the lightcurve rapidly declines).
$r$-band observations are only available after 60~days, and the
search then yields $P_\mathrm{max} = 0.136$ for $N_\mathrm{obs} = 304$.
Both of these $(P_\mathrm{max}, N_\mathrm{obs})$ pairs are periodogram 
outliers. For both bands, we no longer identify
a periodogram peak at 12.4~days when data after 250 days is included.
We obtain higher values of $P_\mathrm{max}$ if we begin the spline fit
later (e.g., after 80 days), when the lightcurve
is not changing as quickly. This suggests that some iterative
or adaptive method could improve our search sensitivity. With
$\lambda = 10$, $P_\mathrm{max}$ is highly sensitive to when the spline fit begins.

We are also well able to recover the 32-day periodicity for
SN~2022esa \citep{SN2022esa}. With the standard search with $\lambda = 1$, we recover
$P_\mathrm{max} = 0.435$ with $N_\mathrm{obs} = 53$ for the $g$ band
and $P_\mathrm{max} = 0.513$ with $N_\mathrm{obs} = 41$ for the $r$ band.
We are unable to recover the periodic signal with the $\lambda=10$ search,
however, as the stiffer spline is unable to adapt to the broad shape
of the underlying SN lightcurve.

The periodicity of 8.4~days for SN~2015ap was only found
by \citet{SN2015ap_Ragosta+} using the Gaussian process approach 
developed by \citet{MindTheGaps}, which
is more robust to the uneven sampling that is often present in observations.
Indeed, we are mostly unable to recover the reported periodicity and 
the periodogram is instead dominated by features introduced by the 
presence of long gaps. 

A few other SNe such as SN~2022mop \citep{SN2022mop_Brennan+}
and SN~2009ip \citep{SN2009ip_1, SN2009ip_2} have also been 
reported to have prominent late-time outbursts with some repetition,
although the available photometric data for these SNe is generally too 
sparse for our method to be applied. 
SN~2022mop 
is of particular relevance: After the initial hydrogen-poor 
SN in 2022, a second, brighter outburst occurred in three years
later with the spectral properties of a type IIn
\citep{SN2022mop_Brennan+}. In the intervening time, there 
were a few indicative fluctuations around a decaying lightcurve.
A potential explanation 
is that the initial SN took place in a binary system 
and resulted in a highly eccentric orbit
before an eventual merger that directly triggered the second SN-like event. 
In this scenario, the fluctuations would be due to accretion onto
the newborn compact object.
If true, then SN~2022mop belongs to
the same family of CC SNe that display photometric evidence
of accretion from a binary companion.


\section{Conclusions}
\label{sec:conclusions}

The discovery of periodic oscillations in SN~2022jli has opened up
the possibility of uncovering more CC SNe with binary interactions.
Indeed, a few other such SNe --- notably SN~2022esa
and SN~2015ap, for which periods have been measured
--- have already been announced. However, searches for candidates
are generally focused on individual, particularly bright SNe, 
whereas a more systematic search is needed to 
understand how detectable the features are in the population. 
We have simulated a set of CC SNe with 
additional periodic and rebrightening features caused by interactions
with a binary companion, assuming a three-day cadence and telescopes
with ZTF-like sensitivities. Overall, the search sensitivity to signals depends
strongly on the number of cycles that are observed; about half of the
simulated parameter space is recoverable with a 200-day observation,
while only 10\% are recoverable with a 75-day observation. At
the same time, the
noise rejection depends directly on the number of observations,
and a 75-day search with 30 days of higher cadence follow-up observations
can achieve the same level of noise rejection as the 200-day observation.
A spline with an intermediate stiffness $\lambda = 1$ can
well recover signals with $T_\mathrm{osc} < 50$~d while a
stiffer spline $\lambda = 10$ is better suited for signals
with $T_\mathrm{osc} > 50$~d. 
Our goal is to design a search
that identifies SN candidates for more precise analysis or deeper follow-up, which
would entail setting a loose cut to find signal candidates (for 
instance, based on an acceptable number of false alarms)
and triggering denser follow-up photometric observations and/or
spectroscopic observations to look for features such as
line velocity shifts.
We recommend using both values of $\lambda$ to ensure sensitivity
over a wider range of binary periods.

Any deviation from a regular cadence affects both
search strategy and results, as
long gaps in the
data --- common for SN observations --- add an additional
timescale and therefore additional structure to the periodogram, 
leading to more false positives. Any search on data
with longer gaps could consider 
alternative methods \citep{SN2015ap_Ragosta+, MindTheGaps},
although these are often more computationally expensive. Instead,
methods based on Gaussian process regression 
are well suited for deeper studies of single candidates that are identified after a periodogram-based
search.

Using our search method, with a three-day cadence,
observation durations of at least
100 days after SN peak are necessary to effectively recover
large portions of the parameter space while maintaining
a manageable number of false alarms. This quantitative
result supports qualitative understanding: There will
always be some delay after the initial supernova for 
sufficient accretion to take place, and a sufficient
number of cycles is necessary to detect any periodicity.
As most SNe are only observed for a month or two, this
suggests that evidence for binary interactions
could have been missed in a large fraction of SNe, 
particularly the ones with longer binary periods.

With LSST as the observatory, the increased depth allows for SNe to be detectable
much longer, although the generally slower cadence would
decrease sensitivity to shorter periods. 
Additionally, if the standard observing
strategy for LSST is to guarantee an observation in \emph{any} band
rather than the same band as before, the single-band
periodogram employed here will suffer due to the large
data gaps. Since we are interested in broadband lightcurve
oscillations, a multiband generalization of the Lomb-Scargle
periodogram \citep{MultibandPeriodograms} would be necessary for LSST data.

Our choices
for the search (in particular, for the spline fit parameters) were
tuned on moderately bright SNe, and a similar 
search on the brightest SNe could need different settings. Indeed,
short-term variability and rapid changes in slope
(both real and instrumental) are more likely
to be observed during the brightest SN lightcurves. 
In addition,
the spline fit uses the flux uncertainties as weights
in finding the optimal interpolation and therefore
relies on these uncertainties being properly calculated, whereas
the brightest SNe can be affected by saturation
and other systematic effects.

A more physically motivated prescription
such as the simulations performed in \citet{Ercolino+2026_arxiv}
would allow for a prediction about the detectability of the population
rather than simply the parameter space. In addition, the set of SN templates
we have used
 --- besides being an observationally biased sample, and
generally limited to SNe detected by older telescopes ---
could already include some lightcurve features caused by interactions
with a binary companion, so that the true rate of
false positives could be lower than determined here.
An improvement would be to generate SN
lightcurves from physics simulations from a targeted subset
such as hydrogen-poor CC SNe. Alternatively,
methods of interpolating or warping lightcurve templates to
produce a more continuous distribution
could reduce the biases in using any particular set of
templates \citep{JakobsStuff}. Either method
would also produce a sufficiently robust training sample for
machine learning methods to identify periodic features.

In this paper, we have demonstrated the proof of concept for a 
simple and computationally inexpensive search for supernovae exhibiting
multiple periodic oscillations in their lightcurves, that can be run
in realtime on data from optical surveys such as ZTF and LSST. 
This is the first step to characterizing this behavior on a 
population level, to better understand the evolution of massive
stars in binaries.

\begin{acknowledgements}
    The authors thank Steve Schulze for very helpful discussions. This
    manuscript made use of \texttt{numpy} \citep{numpy}, 
    \texttt{scipy} \citep{scipy}, and
    \texttt{matplotlib} \citep{matplotlib}.
\end{acknowledgements}

\bibliographystyle{aa} 
\bibliography{refs} 

\begin{appendix}

\section{Supplementary information}

\subsection{Conversions}\label{sec:basics}

We would like to make a few conversions explicit that are
obvious to optical astronomers --- including at least one of the
authors --- but proved difficult for another of the authors to 
track down and understand. We do this with the aim of improving
communication between scientists from different fields, who
often use different vocabulary and units to discuss the same
properties.

For converting between magnitude and flux in our simulations, 
we assume a zeropoint ZP of 25, appropriate for ZTF. The
explicit conversion between flux $F$ and magnitude $m$ with
$\mathrm{ZP}=25$ is
\begin{align}
    F = 10^{-(m - 25)/2.5} \quad\longleftrightarrow\quad m = 25 - 2.5 \log_{10} F.
\end{align}\label{eq:fluxmag}
$F$ is therefore a scaled flux, without physical units. It
only has meaning with a known zeropoint, or for comparisons
with other scaled fluxes using the same zeropoint.

The ZTF IPAC server generally returns fluxes in units of
digital number DN. The conversion between flux in units of DN
and flux in units of Jy (or, more precisely, the specific flux density
$F_\nu$) is (e.g, Eq.~9 of \citet{Fconversion})
\begin{align}
    F_\mathrm{[Jy]} = F_\mathrm{[DN]} \times 10^{(8.9 - \mathrm{ZPmag_{DN}})/2.5}
\end{align}\label{eq:fluxes}
where $F_\mathrm{[DN]}$ is the quantity \texttt{forcediffimflux}
and $\mathrm{ZPmag_{DN}}$ is \texttt{zpdiff}.

\subsection{Sky noise for simulations}\label{sec:skynoise}

\begin{figure}
    \centering
    \includegraphics[width=0.9\hsize]{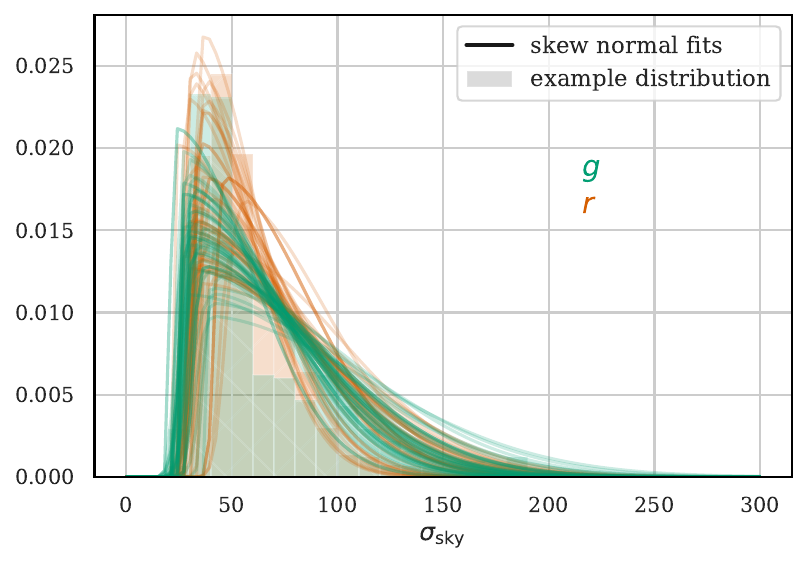}
    \caption{Skynoise $\sigma_\mathrm{sky}$ as an input to the
    SN simulations, derived for the $g$ and $r$ bands separately
    from ZTF data. For all of our simulated SNe, we 
    drew $\delta_\mathrm{sky}$ from a skew normal fit to all visits
    pertaining to a specific sky position in a given field. Here,
    we show 50 sample fits as well an example distribution for
    the $g$ and $r$ bands separately.}
    \label{fig:skynoise}
\end{figure}

Sky noise is an input for \texttt{realize\_lcs()} that quantifies the
background noise. In \texttt{realize\_lcs()} the skynoise 
$\sigma_\mathrm{sky}$ combines with the the uncertainty 
the photon counts in quadrature:
\begin{align}
    \sigma_F = \sqrt{\sigma_\mathrm{sky}^2 + F}
\end{align}\label{eq:skynoise}
where $F$ is the photon counts flux and $\sigma_F$ is the 
total uncertainty on the flux. $\sigma_\mathrm{sky}$ describes
the total noise due to observing conditions; for a given sky position,
this includes both constant effects (such as the general sky brightness
in that region) as well as temporary effects (such as the atmospheric
conditions during a particular observation). We determined
$\sigma_\mathrm{sky}$ from the IPAC forced photometry server 
\citep{IPAC1, IPAC2} for the 212 ZTF SNe discussed in Sec.~\ref{sec:ztf},
and calculated $\sigma_\mathrm{sky}$ using
\begin{align}
    \sigma_\mathrm{sky} = \frac{10^{0.4 \times (\texttt{zpdiff} - \texttt{diffmaglim})}}{5}
\end{align}
where \texttt{zpdiff} is the photometric zeropoint and \texttt{diffmaglim}
the magnitude limit for the difference image, as in Eq.~1 of 
\citet{2025A&A...694A...3A}.
For each band and field, we fit the 
distribution of $\sigma_\mathrm{sky}$ with a skew normal distribution
for a total of 245 $g$-band fits and 247 $r$-band fits.
(This is more than the number of SNe because some SNe
appeared in multiple fields.) A sample of 50
random fits as well as an example distribution is shown in 
Fig.~\ref{fig:skynoise} for the $g$ and $r$ bands separately. In general,
the $\sigma_\mathrm{sky}$ distributions in $g$ peak at smaller values
than in $r$ but have longer tails.

For any SN simulation in a given band, we drew the skynoise 
for each observation from a randomly chosen skew normal fit for 
that band. Since we ran our searches on the $g$- and $r$-band simulations
independently, we did not require that the fits be drawn from the same
field.

Note that Eq.~\ref{eq:skynoise} can be inverted to solve for
the flux $F$ when $\sigma_\mathrm{sky}$ and $\sigma_F$ are known.
This is important because \texttt{sncosmo} returns flux points
$F_\mathrm{SN}$ that have already been randomly redrawn from a one-dimensional
Gaussian with a width of $\sigma_{F_{SN}}$, whereas we want the option
to add rebrightening and oscillation features to the template flux
before calculating uncertainty. Therefore, for our simulations, we inverted
Eq.~\ref{eq:skynoise} to calculate the interpolated template flux
$F_\mathrm{SN}$ given the reported $\sigma_F$ and the
input $\sigma_\mathrm{sky}$, and then
added the flux due to the rebrightening $F_\mathrm{rebr}$ and
oscillation features $F_\mathrm{osc}$ when necessary so that
$F = F_\mathrm{SN} + F_\mathrm{rebr} + F_\mathrm{osc}$. We then 
calculated the overall uncertainty $\sigma_F$ with the new 
total flux $F$, and redrew $F$ from the Gaussian with the
width $\sigma_F$.

\subsection{More explorations of recoverable signals}
\label{sec:apprecovery}

\begin{figure*}[ht!]
    \centering
    \includegraphics[width=\textwidth]{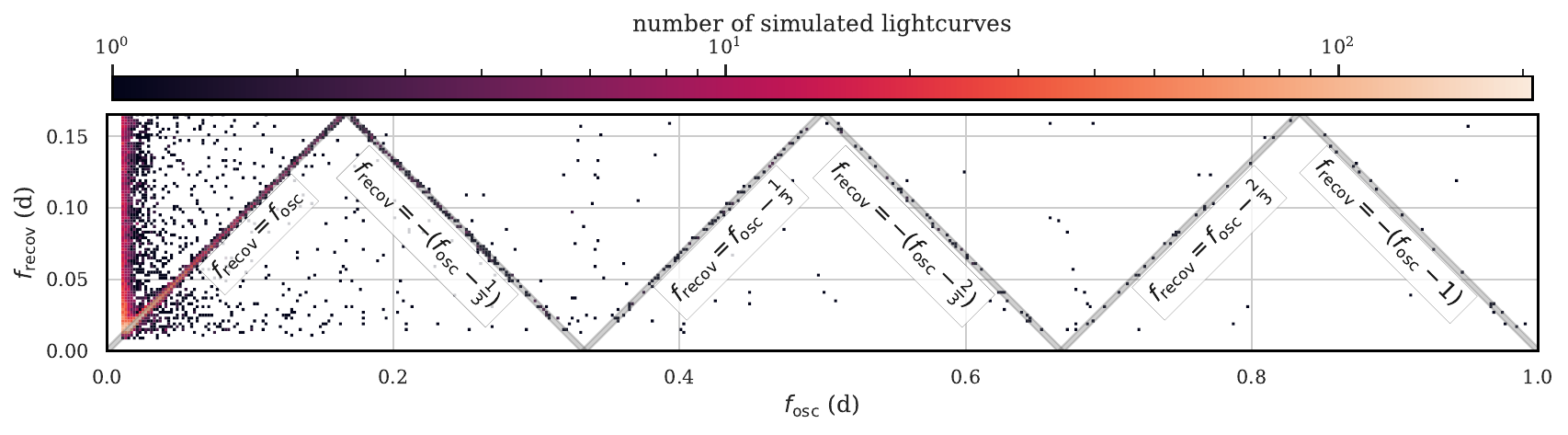}
    \caption{2D histogram showing $f_\mathrm{recov}$ across
    the range of $f_\mathrm{osc}$ for the basic search with
    $\lambda=1$ starting at $t_0 + 20$~d. 
    In addition to the set of signals with 
    $f_\mathrm{recov} \approx f_\mathrm{osc}$, 
    the sets of aliased signals that correspond to
    Eq.~\ref{eq:aliases_f} with $n < 0$ are annotated. These are
    equivalent to the ones highlighted in Eq.~\ref{eq:aliases_T1}.
    The gray shaded regions around the aliased signal sets depict the
    $f_\mathrm{recov} = f_\mathrm{osc} \pm 0.0021$ definition.}
    \label{fig:example_lam1spl20_2Dhist}
\end{figure*}

Our definition for a recoverable signal is a criterion on
$f_\mathrm{LS}$. Fig.~\ref{fig:example_lam1spl20_2Dhist}
shows a 2D histogram of the frequencies $f_\mathrm{recov}$ associated
with the periodogram peaks $P_\mathrm{max}$, compared to the simulated
frequency $f_\mathrm{osc}$, for the standard search ($\lambda = 1$, 
starting at $t_0 + 20$~d). The recovery condition of both the
signals around $f_\mathrm{recov} = f_\mathrm{osc}$ and
the aliased signals described in Eq.~\ref{eq:aliases_T1} 
are better defined in frequency space, as $f_\mathrm{recov} = f_\mathrm{osc} \pm 0.0021$~d$^{-1}$ (or, for the aliased signals, around an alias of $f_\mathrm{osc}$);
i.e., the recovered frequency is within 4.2 LS gridpoints $\Delta f_\mathrm{LS}$ of the simulated frequency. The choice of 4.2 gridpoints
well captures the spread around $T_\mathrm{recov} = T_\mathrm{osc}$
for large $T_\mathrm{osc}$, as seen in Fig.~\ref{fig:scatter_6days},
without being too restrictive. The choice of 4.2 as
opposed to exactly 4 gridpoints is to take into account differences
due to rounding.

\begin{figure}[ht!]
    \centering
    \includegraphics[width=\hsize]{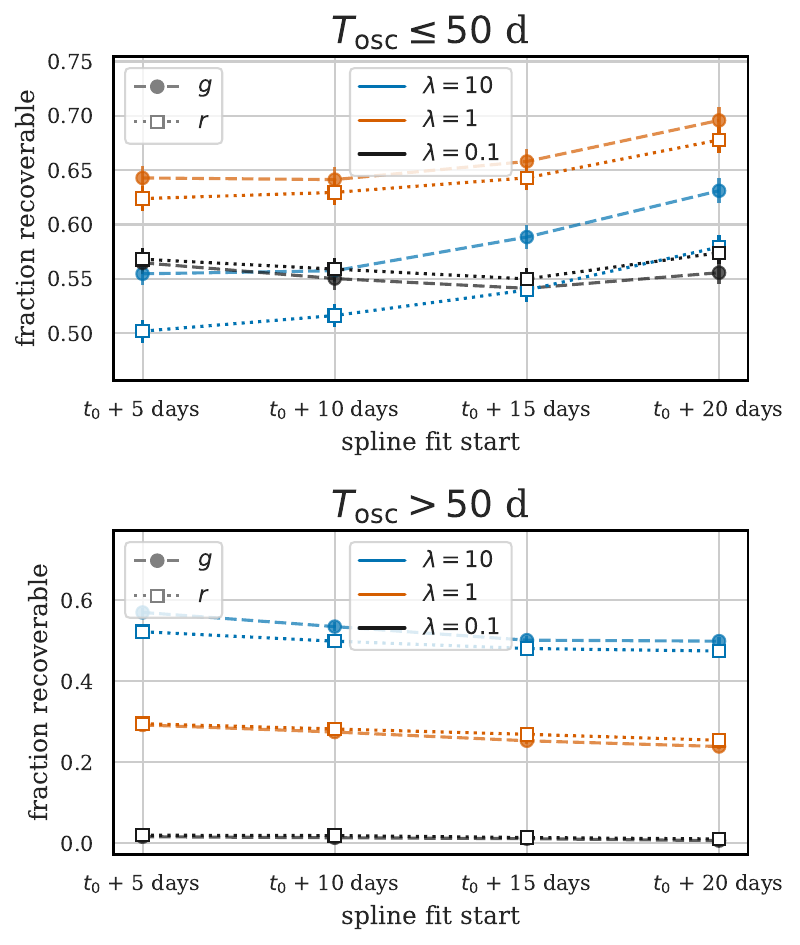}
    \caption{Comparing fraction of recoverable signals for
    $g$- (closed circles) and $r$-band (open squares) 
    simulations, across different values of
    $\lambda$ and the spline fit start day, for $T_\mathrm{osc} \leq
    50$~d (top) and $T_\mathrm{osc} > 50$~d (bottom). Fewer signals
    are recovered with the $r$-band simulations, especially
    for shorter period signals and using $\lambda = 10$.}
    \label{fig:splineFits_bands}
\end{figure}

The recoverability for the $r$ and $g$ bands are compared in
Fig.~\ref{fig:splineFits_bands}. The fraction recovered in
the $r$ band is consistently worse than in the $g$ band,
with the difference being at most 5\% (for $\lambda = 10$ 
and short period signals). This suggests that the $r$-band
lightcurves vary more rapidly for the
templates we are using, which obscures the underlying oscillations.

\begin{figure*}[ht!]
    \centering
    \includegraphics[width=\textwidth]{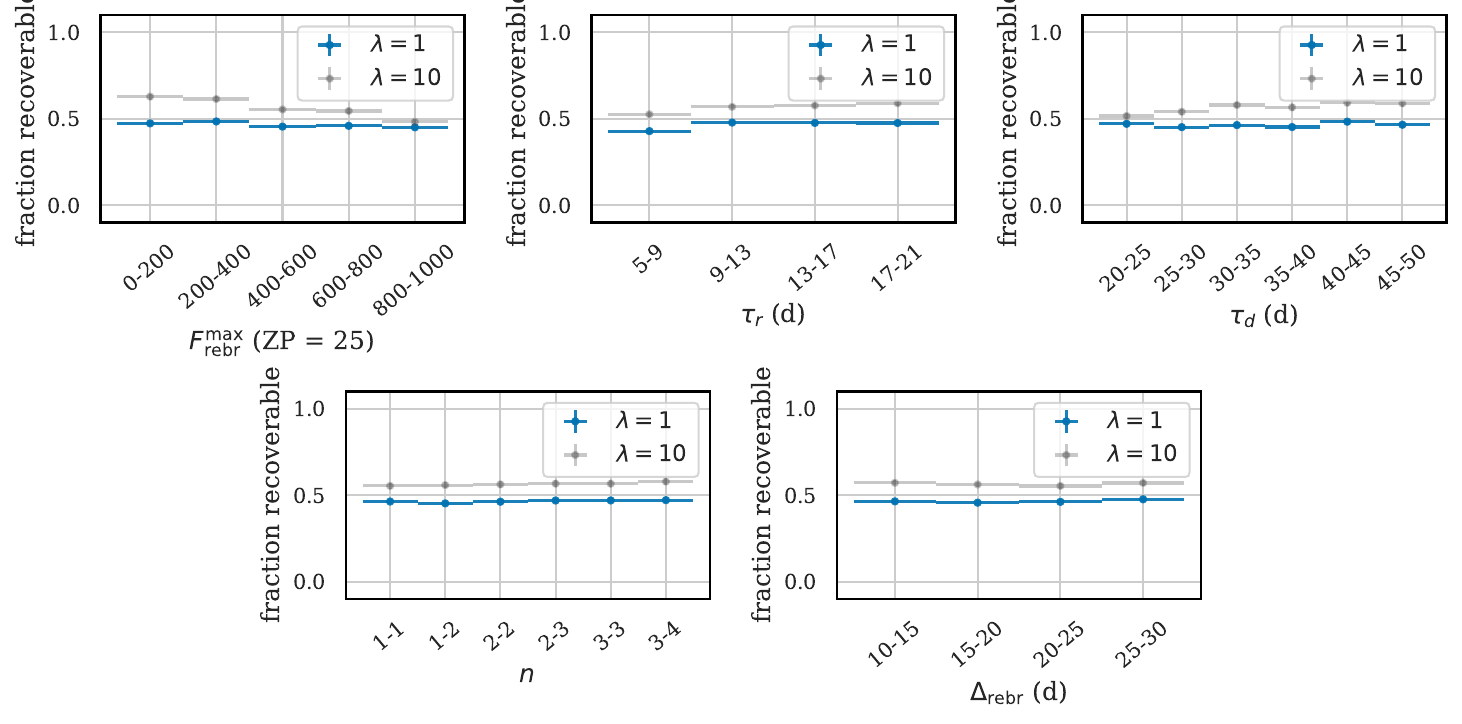}
    \caption{The difference in recovered versus true
    frequency as a
    function of the rebrightening parameters for the simulated SNe. 
    None of the five rebrightening parameters has a strong effect on the signal recovery for $\lambda=1$.}
    \label{fig:tripleScatter_rebrParams}
\end{figure*}

Both the rebrightening and oscillating features of SN~2022jli are attributed
to accretion of the binary companion, although it is unclear how universal
they are. However, for our standard search, the rebrightening features do
not have any effect on how $\left| f_\mathrm{recov} - f_\mathrm{osc} \right|$
is distributed
(Fig.~\ref{fig:tripleScatter_rebrParams}),
suggesting that our search with $\lambda=1$ is robust to
the kinds of rebrightening features found in SN~2022jli.

\subsection{Non-symmetric oscillations}
\label{app:sawtooth}

\begin{figure}[ht!]
    \centering
    \includegraphics[width=0.9\hsize]{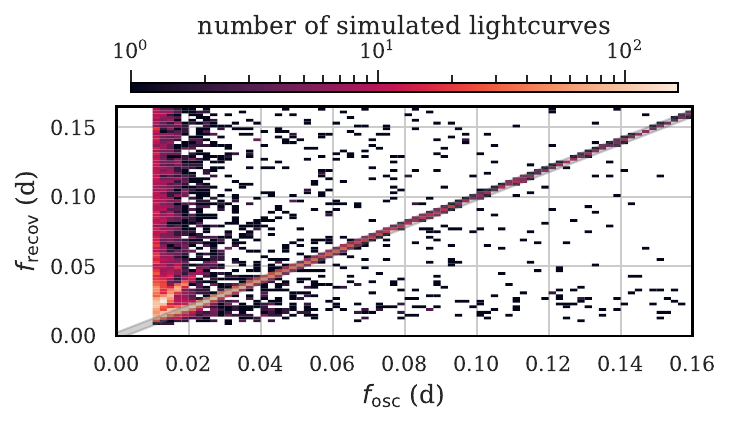}
    \caption{Like Fig.~\ref{fig:example_lam1spl20_2Dhist} but for the skewed
    sawtooth oscillations rather than symmetric sinusoids. The sawtooth
    functions have a ratio of rise time to fall time of 1:4 (i.e., they take
    four times as long to decay as to rise). The $x$ axis
    is limited to $[0, 0.16]$ in order to highlight the new feature:
    besides the overdensity around $f_\mathrm{recov} = f_\mathrm{osc}$,
    there is an additional overdensity at $f_\mathrm{recov} = 
    2f_\mathrm{osc}$,
    particularly for longer period signals ($f_\mathrm{osc} < 0.03$~d),
    leading to a 5\% reduction in the number of
    signals recoverable around $f_\mathrm{osc}$ or an alias.}
    \label{fig:example_lam1_saw_2Dhist}
\end{figure}


While the oscillations for SN~2022esa were symmetric \citep{SN2022esa},
the oscillations for SN~2022jli were skewed with shorter
rise times than fall times \citep{SN2022jli_Chen+}.
If the periodic oscillations are skewed, this means
that some power will instead be present at higher order frequencies,
so that our sensitivity to real signals would decrease. To test this,
we resimulated the 10 000 SN lightcurves with the periodic oscillations
being skewed sawtooth functions rather than symmetric sinusoids, 
keeping all other parameters such as the period the same. For the
sawtooth functions, we set the fall time to be four times as long as the
rise time, similar to what was seen for SN~2022jli \citep{SN2022jli_Chen+}.

As Fig.~\ref{fig:example_lam1_saw_2Dhist} shows, when compared to the
equivalent search on symmetric sinusoidal oscillations as depicted in
Fig.~\ref{fig:example_lam1spl20_2Dhist}, there is an additional overdensity
of signals recovered at $f_\mathrm{recov} = 2f_\mathrm{osc}$. This is 
especially true for the longer period signals (the left side of the figure). 
Overall, this reduces the number of recoverable signals by
5\% for both the $\lambda=1$ and $\lambda=10$ searches 
as there is less power in the periodogram
at frequencies around $f_\mathrm{osc}$ and its aliases. If we additionally
consider the first higher order mode by including signals recovered at
$f_\mathrm{recov} = 2f_\mathrm{osc}$ and its aliases, the missing
5\% of signals are mostly recovered but at much lower $P_\mathrm{max}$
values.

\subsection{Search grids with smaller $T^\mathrm{min}_\mathrm{LS}$}
\label{sec:shorter}

\begin{figure*}[ht!]
    \centering
    \includegraphics[width=\textwidth]{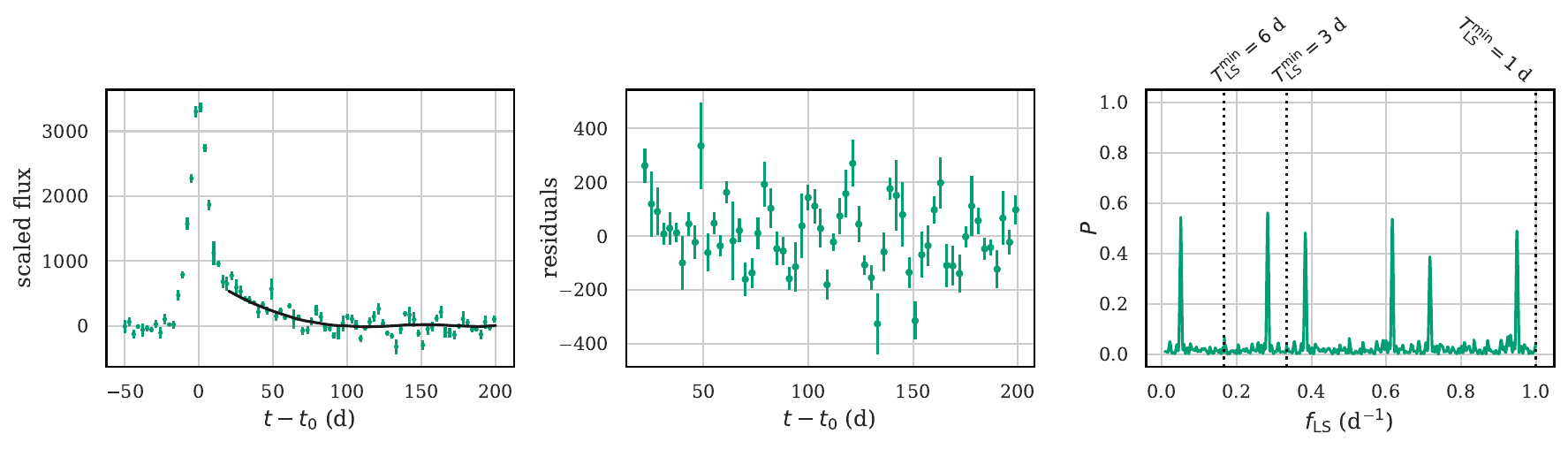}
    \caption{Lightcurve and spline fit (left), residuals (middle), and
    periodogram (right) for an example simulation with $T_\mathrm{osc} = 20$~d
    and a three-day cadence. The periodogram shows the reflected regions 
    beyond $T_\mathrm{LS}^\mathrm{min} < 6$~d, including the aliased
    (i.e., reflected) versions of the signals at higher frequencies. Fluctuations
    determine the relative heights of the peaks, so that $f_\mathrm{recov}$
    could be at either the main frequency $f_\mathrm{osc}$ or one of the
    aliased frequencies.}
    \label{fig:example_aliases}
\end{figure*}

When using our standard periodogram grid with $T^\mathrm{LS}_\mathrm{min} = 6$~d,
we can stillrecover signals with
$T_\mathrm{osc} < 6$~d via their longer period aliases.
Alternatively, we can also use search grids
that extend down to shorter periods to potentially increase
our sensitivity to shorter oscillations, and accept that some longer
period signals will be recovered by their shorter period aliases:
\begin{align}\label{eq:aliases_T2}
    T_\mathrm{recov} = \left(\frac{1}{T_\mathrm{osc}} - n \delta f_\mathrm{obs}\right)^{-1}
\end{align}
(e.g., Eq.~45 in \citet{LombScargle_VanderPlas}\footnote{See footnote associated
with Eq.~\ref{eq:aliases_T2} on the choice of sign convention.}). An example of this
concept is shown in Fig.~\ref{fig:example_aliases}.

\begin{figure*}[ht!]
    \centering
    \includegraphics[width=\textwidth]{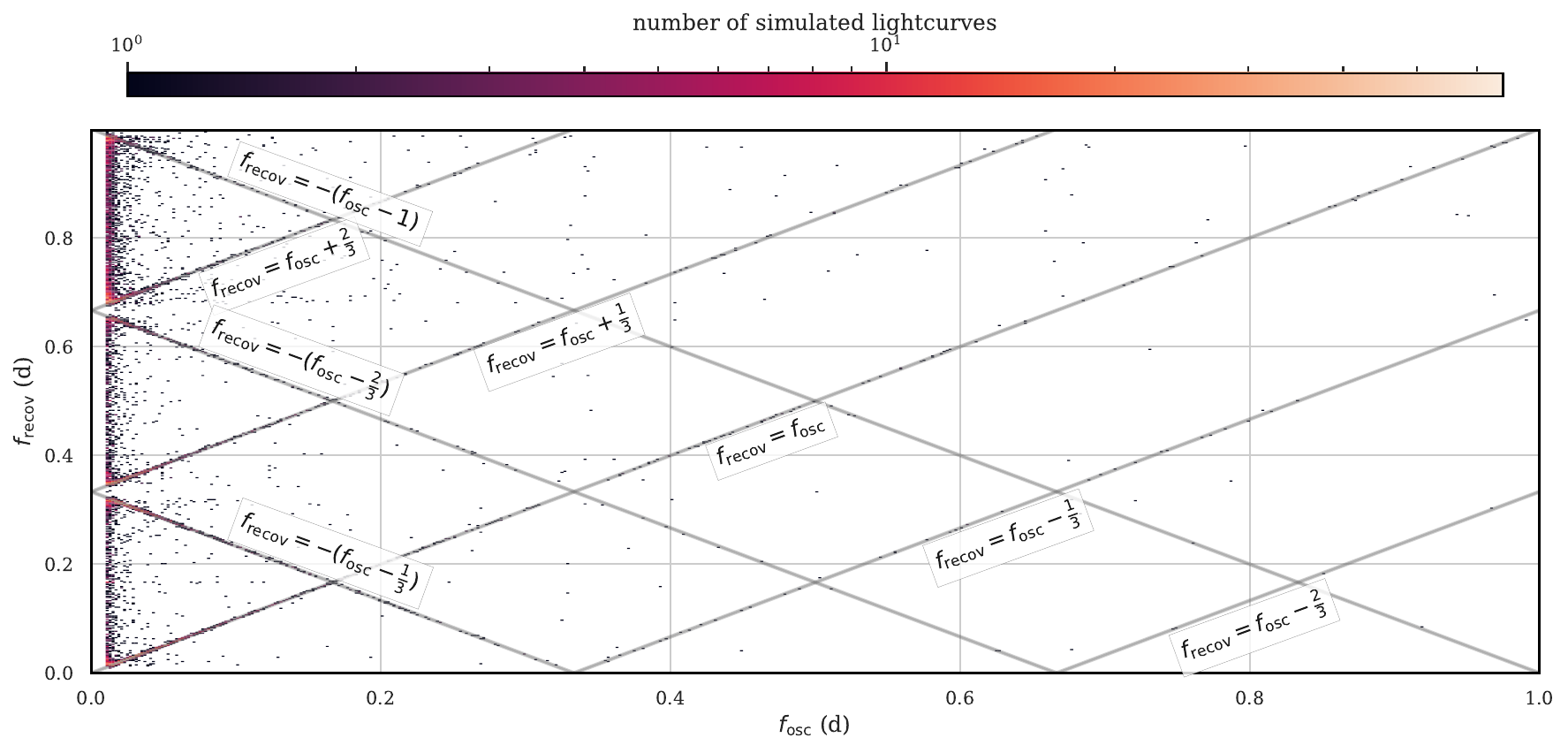}
    \caption{Same as Fig.~\ref{fig:example_lam1spl20_2Dhist} but using the
    periodogram grid with $T_\mathrm{LS}^\mathrm{min} = 1$~d. (The figure is
    compressed in the vertical direction so the lines are at different angles
    compared to the previous figure.) The search grid is six times as large as
    the one with $T_\mathrm{LS}^\mathrm{min} = 6$~d, and the plot itself can
    be seen as the one in Fig.~\ref{fig:example_lam1spl20_2Dhist} reflected
    six times over the vertical axis. This shows the additional aliases
    that appear when using a grid with $T_\mathrm{LS}^\mathrm{min}$ less than
    twice the cadence.}
    \label{fig:example_lam1spl20_LSgrid3_2Dhist}
\end{figure*}

\begin{figure}
    \centering
    \includegraphics[width=0.9\hsize]{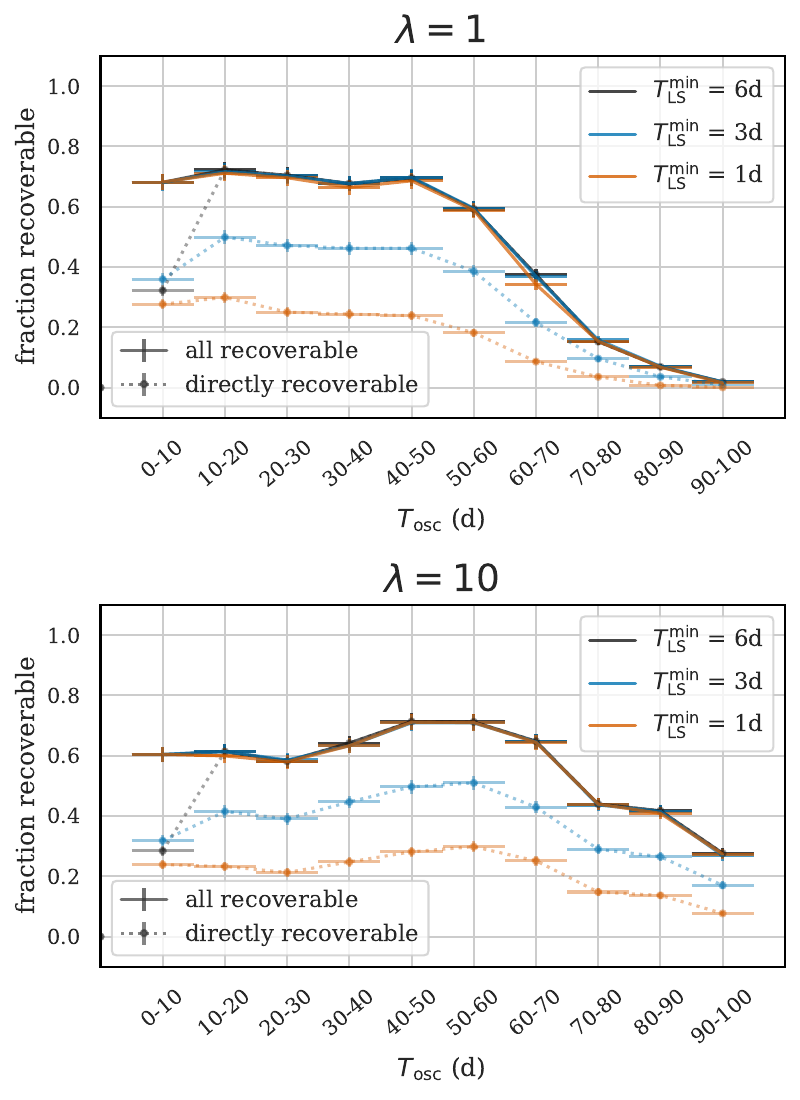}
    \caption{Signal recoverability as a function of $T_\mathrm{osc}$ for 
    periodogram grids with values of $T_\mathrm{LS}^\mathrm{min}$,
    for the $\lambda = 1$ (left) and $\lambda = 10$ (right) spline fits. 
    There is no major difference between the total number of recoverable
    signals for the three choices of periodogram grid, although
    larger fractions are recoverable only at their aliased frequencies
    with the smaller values of $T_\mathrm{LS}^\mathrm{min}$.
    Vertical error bars show counting 
    uncertainties.}
    \label{fig:numPassed_Tmins}
\end{figure}

To explore this, we also tried search grids with  $T_\mathrm{LS}^\mathrm{min}$
of three days and one days. The
search results for $T^\mathrm{LS}_\mathrm{min} = 1$~d are shown
in Fig.~\ref{fig:example_lam1spl20_LSgrid3_2Dhist}, to help visualize
the signals that are recovered at aliased frequencies.
Overall, the three periodogram grids recovered similar numbers
of signals (Fig.~\ref{fig:numPassed_Tmins})
for both $\lambda = 1$ and $\lambda = 10$,
although more of these are recovered at their aliased frequencies
for $T_\mathrm{LS}^\mathrm{min} = 3$~d and 1~d. Interestingly, the
search grids with smaller $T_\mathrm{LS}^\mathrm{min}$ do not 
recover any additional short period signals compared to the 
grid with $T_\mathrm{LS}^\mathrm{min} = 6$~d, indicating that the
short period signals are efficiently recovered at their aliased frequencies.
We therefore find no benefit in using search grids with smaller values
of $T_\mathrm{LS}^\mathrm{min}$.

\subsection{Using only stripped envelope SN templates}
\label{app:SE}

In our main search, we have included all CC SN templates available in
\texttt{sncosmo}. However, this also encompasses many SNe that are
classified as non-stripped envelope SNe, including some with less well behaved
lightcurves, and in principle using only stripped envelope SN templates
could produce a better and more targeted result. To explore this,
we reran the $\lambda = 1$ and $\lambda=10$ searches on a new set of
10 000 SNe simulations that only included the Ib, Ic, and IIb templates,
as well as their subtypes such as Ic-BL. The searches started 
$t_0 + 20$~d and ran on the simulated lightcurves with the standard three-day 
cadence and 200-day observation period, using the periodogram grid with
$T_\mathrm{LS}^\mathrm{min} = 6$~d.

\begin{figure}[ht!]
    \centering
    \includegraphics[width=0.9\hsize]{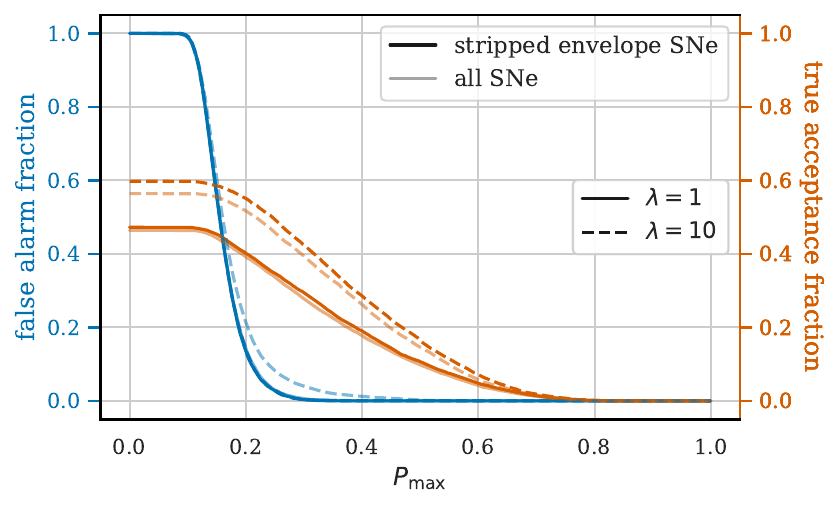}
    \caption{Comparing the false alarm and true acceptance fractions at different 
    $P_\mathrm{max}$ thresholds when simulating the SNe only from the templates of
    stripped envelope SNe. The $\lambda=1$ (solid curves) and $\lambda=10$ 
    (dashed curves) searches both start at $t_0 + 20$~d. The equivalent quantities
    for the simulations that include all SN templates are plotted as the semitransparent
    curves. Using a more limited set of templates does not result in a big
    difference in either the signal recovery or the noise rejection.}
    \label{fig:FATA_SE}
\end{figure}

We find that, when only using the stripped envelope SN templates, the
recovered SNe increase by less than a percent for the $\lambda = 1$ search
and 3.5\% for the $\lambda = 10$ search. Figure~\ref{fig:FATA_SE} shows
this comparison when using the more limited set of templates versus the
whole set of core collapse SN templaets. The false alarm fractions
are also not strongly affected by these different set of templates.

\begin{figure}[ht!]
    \centering
    \includegraphics[width=0.9\hsize]{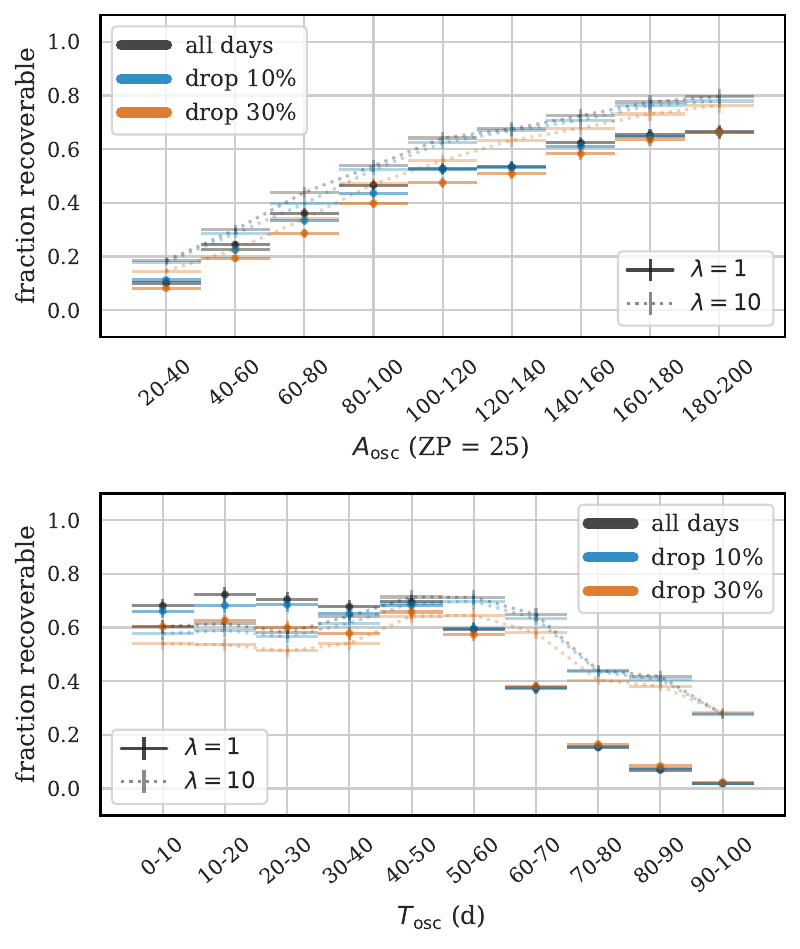}
    \caption{The effects of randomly dropping 10\% of observations
    is a minimal loss of 1\% of detectable signals, while randomly 
    dropping 30\% of observations reduces the detectable percentage
    by 5\%.}
    \label{fig:dropdays}
\end{figure}

\begin{figure}
    \centering
    \includegraphics[width=0.9\hsize]{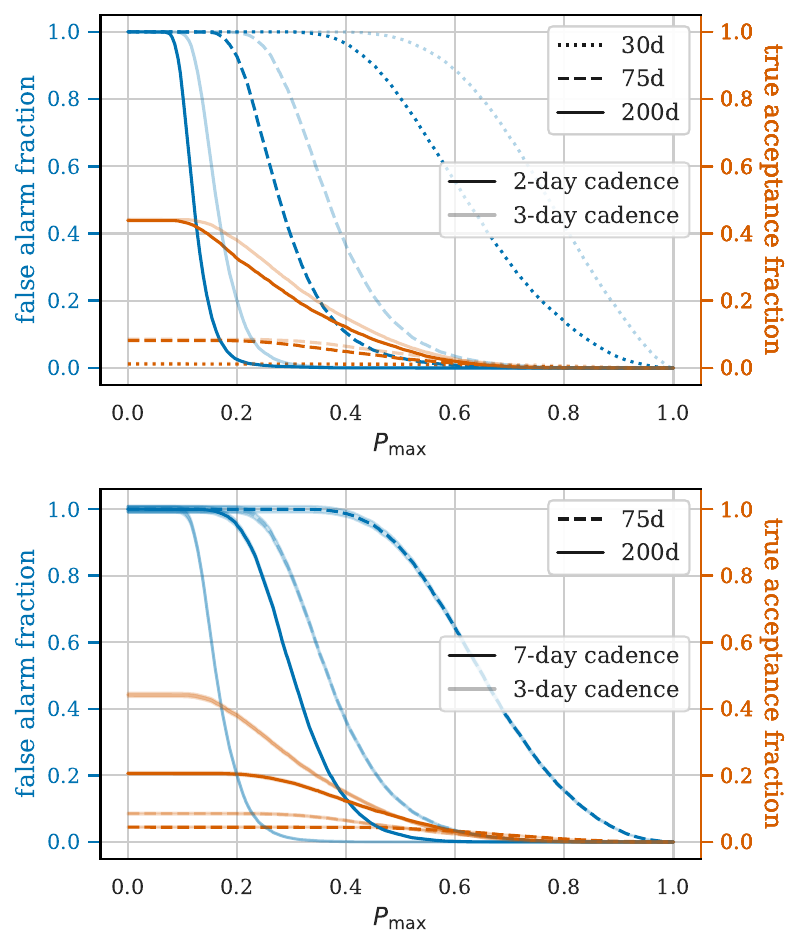}
    \caption{Similar to Fig.~\ref{fig:FATA_threeDurations} but with the two- (top) or seven-day
    cadence (bottom) observations, compared to our standard three-day cadence assumption.}
    \label{fig:FATA_cadences}
\end{figure}



\subsection{Properties of the ZTF-detected SNe}
\label{app:ztf}

\begin{figure*}
    \centering
    \includegraphics[width=0.65\textwidth]{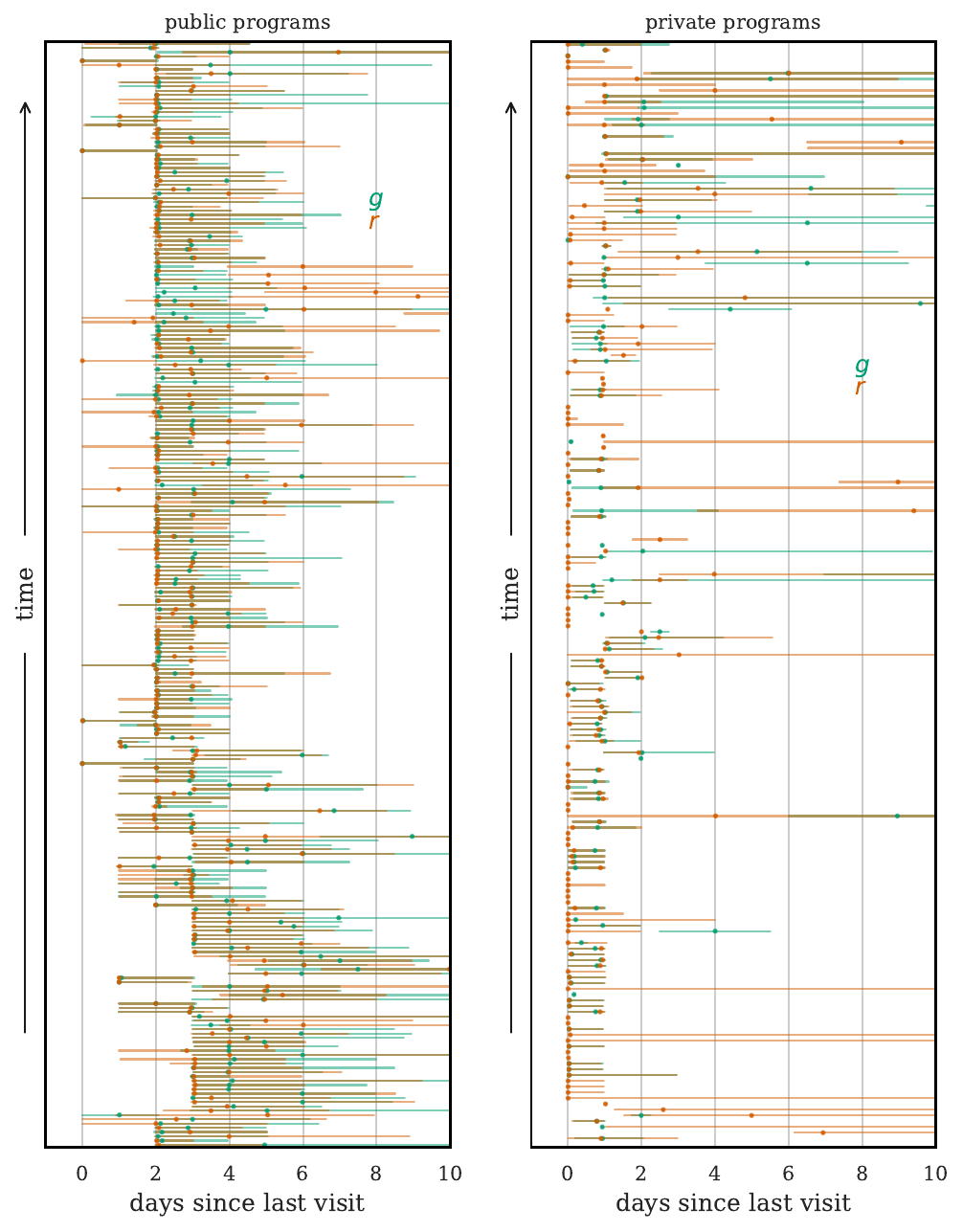}
    \caption{The mean (circle) and standard deviation (bars) in the days
    since the previous visit, for all SNe in our moderately bright ZTF sample,
    split into the $g$ (green) and $r$ (red) bands. The vertical axis shows
    increasing time, from bottom to top, and each point corresponds to a single
    field in a given SN; since some SNe were present in multiple fields, there
    are more points than there are SNe.
    The standard three- and two-day cadences of the public surveys (left) are
    visible, while the private programs (right) often include multiple observations
    a night. For the data plotted here, fields with IDs >1000 (i.e.,
    the secondary field grid)
    were removed but no other cleaning was performed. Some SNe included
    visits from both public and private programs.}
    \label{fig:ZTF_mid_cadences}
\end{figure*}

The ZTF data we used included visits from both the public and private programs.
While the public programs had standard three- and two-day cadences, the private
program included both higher (i.e., multiple observations per night) and lower
cadences. The mean and standard deviation in the days since the last visit for
any given field are plotted in Fig.~\ref{fig:ZTF_mid_cadences} for the
moderately bright sample, and the difference in cadence between the public
and private surveys can clearly be seen, as well as the switch from the
three- to two-day public surveys.


For the moderate ZTF sample, we drew the simulated SNe from a redshift
range $[0.03, 0.06]$ to match the range of the real SNe, and the scaling parameter
from the range $[10^{-16}, 5\times10^{-15}]$ to produce a similar set of peak
brightnesses as the real sample. 
For the
bright and nearby sample, we used a redshift range of $[0.001, 0.03]$ and a
narrow scaling parameter range $[2\times10^{-14}, 3\times10^{-14}]$, again to match the peak brightness
distribution of the real sample. 


\begin{figure*}
    \centering
    \includegraphics[width=0.9\textwidth]{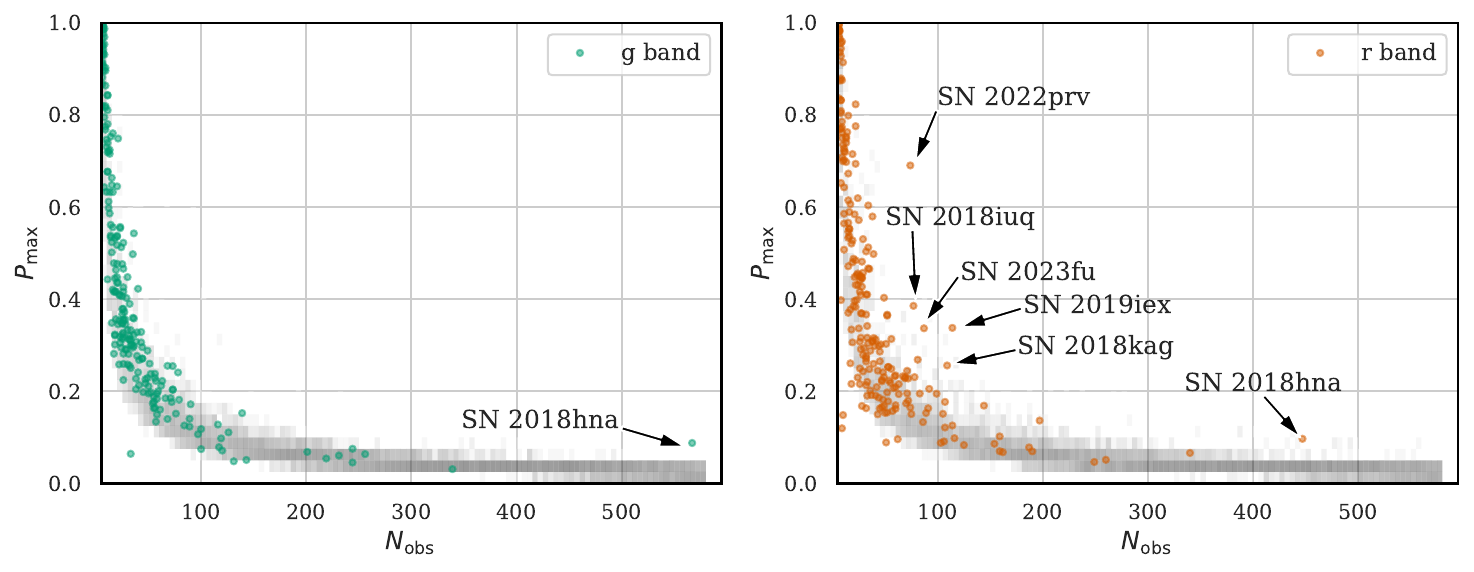}
        \caption{$P_\mathrm{max}$ vs. the number of data points 
        included in the spline fit $N_\mathrm{obs}$
        for the $g$ and $r$ bands individually, for the nearby
        and very bright ZTF SN sample, with the results
        from 10~000 simulations shown in the gray 2D histogram in the
        background (logarithmic scaling). The outlier SNe are
        labeled; all can be discarded after visual 
        inspection of the lightcurves and spline fits. The only
        SN that appears in both the $g$- and $r$-band plots,
        SN~2018hna, was covered by multiple fields explaining
        both the large $N_\mathrm{obs}$ as well as the $P_\mathrm{max}$ outlier due to the resultant poor spline fit.}
    \label{fig:ZTF_bright_Pmaxs}
\end{figure*}

\begin{figure*}[ht!]
    \centering
    \includegraphics[width=\textwidth]{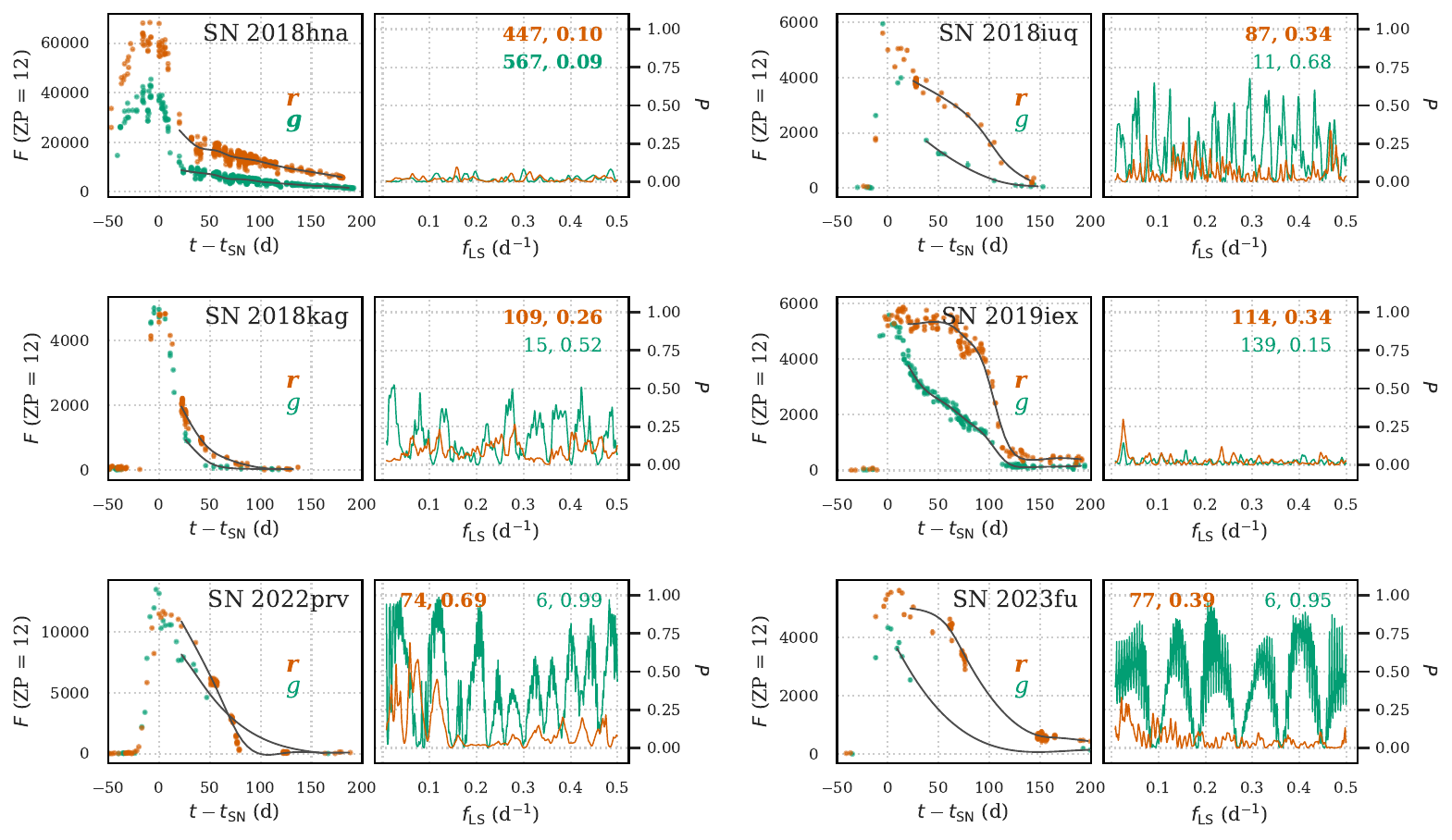}
    \caption{Examples of SN lightcurves, spline fits, and
    periodograms for the nearby ZTF sample. These six periodogram
    outliers can all be excluded as real signal candidates. 
    SN~2018hna, the only $g$-band outlier, appeared as a periodogram
    outlier in both the $g$ and $r$ bands due to its 
    position being present in
    multiple fields. Of the others, SNs~2019iex and 2022prv
    demonstrate the limitation of the spline fit, which was
    tuned to a moderate bright sample and cannot accommodate
    the larger changes in slope and short-term variability (both
    real and instrumental) more
    frequently present in brighter lightcurves.}
    \label{fig:ZTF_bright_examples}
\end{figure*}

\end{appendix}
\end{document}